\documentclass[10pt,journal,compsoc]{IEEEtran}
\IEEEoverridecommandlockouts

\ifCLASSINFOpdf
   \usepackage[pdftex]{graphicx}
\else
   \usepackage[dvips]{graphicx}
\fi

\ifCLASSOPTIONcompsoc
	 \usepackage[caption=false,font=footnotesize,labelfont=sf,textfont=sf]{subfig}
\else
  \usepackage[caption=false,font=footnotesize]{subfig}
\fi

\usepackage{cite}
\usepackage[cmex10]{amsmath}

\usepackage{bm}
\usepackage{amsfonts}
\usepackage{array}
\usepackage{mdwmath}
\usepackage{mdwtab}

\usepackage{fixltx2e}
\usepackage{url}
\usepackage{tabularx}
\usepackage{multirow}
\usepackage{booktabs}
\usepackage{threeparttable}
\usepackage{setspace}
\usepackage{makecell}

\usepackage{enumitem}

\usepackage[ruled,vlined]{algorithm2e}

\usepackage{supertabular}

\usepackage{amsmath,amssymb}

\usepackage[fleqn,tbtags]{mathtools}

\usepackage{lstlinebgrd}

\usepackage{listings}
\lstloadaspects{formats}

\lstdefineformat{C}
{
  \{=\newline\string\newline\indent,%
  \}=\newline\noindent\string\newline,%
  ;=[\ ]\string\space,%
}

\usepackage{color}
\definecolor{dkgreen}{rgb}{0,0.6,0}
\definecolor{gray}{rgb}{0.5,0.5,0.5}
\definecolor{mauve}{rgb}{0.58,0,0.82}
\usepackage{lipsum}

\newtheorem{definition}{Definition}

\begin{document}
\title{ARJA: Automated Repair of Java Programs via Multi-Objective Genetic Programming}
\author{Yuan Yuan and Wolfgang Banzhaf
\thanks{Y. Yuan and W. Banzhaf are with the Department of Computer Science and Engineering, Michigan State University, East Lansing, MI 48824 USA (e-mail: yyuan@msu.edu;  banzhafw@msu.edu).}
\thanks{Manuscript received xxx; revised yyy.}
}

\markboth{Journal of \LaTeX\ Class Files,~Vol.~14, No.~8, August~2015}%
{Shell \MakeLowercase{\textit{et al.}}: Bare Demo of IEEEtran.cls for Computer Society Journals}

%

\IEEEtitleabstractindextext{%
\begin{abstract}
Automated program repair is the problem of automatically fixing bugs in programs in order to significantly reduce the
debugging costs and improve the software quality. To address this problem, test-suite based repair techniques 
regard a given test suite as an oracle and modify the input buggy program to make the whole test suite pass.
GenProg is well recognized as a prominent repair approach of this kind, which uses genetic programming (GP) to rearrange the statements already extant in the buggy program. 
However, recent empirical studies show that the performance of GenProg is not satisfactory, particularly for Java. 
In this paper, we propose ARJA, a new GP based repair approach for automated repair of Java programs.  
To be specific, we present a novel lower-granularity patch representation that properly decouples
the search subspaces of likely-buggy locations, operation types and potential fix ingredients, enabling
GP to explore the search space more effectively. Based on this new representation, we formulate 
automated program repair as a multi-objective search problem and use NSGA-II to look for 
simpler repairs. To reduce the computational effort and search space, we introduce a test filtering procedure
that can speed up the fitness evaluation of GP and three types of rules that can be applied to
avoid unnecessary manipulations of the code. Moreover, we also propose a type matching strategy that can create
new potential fix ingredients by exploiting the syntactic patterns of the existing statements. 
We conduct a large-scale empirical evaluation of ARJA along with its variants on both seeded bugs and real-world bugs
in comparison with several state-of-the-art repair approaches. Our results verify the 
effectiveness and efficiency of the search mechanisms employed in ARJA and also 
show its superiority over the other approaches. In particular, compared to jGenProg (an implementation of GenProg for Java),
an ARJA version fully following the redundancy assumption can generate a test-suite adequate patch for 
more than twice the number of bugs (from 27 to 59), and a correct patch for nearly four times of the number (from 5 to 18), on 224 real-world bugs considered in Defects4J. 
Furthermore, ARJA is able to correctly fix several real multi-location bugs that are hard to be repaired by most of the existing repair approaches.

\end{abstract}

}

\maketitle

\IEEEdisplaynontitleabstractindextext
\IEEEpeerreviewmaketitle

\section{Introduction}
\label{sec-Introduction}

\IEEEPARstart{A}{utomated} program repair \cite{weimer2010automatic,monperrus2017automatic, gazzola2017automatic} 
aims to automatically fix bugs in software. This research field has recently stirred great interest in the software engineering community
since it tries to address a very practical and important problem. 
Automatic repair techniques generally
depend on an oracle which can consist of a test suite \cite{le2012genprog}, pre-post conditions \cite{pei2014automated} or an abstract behavioral model \cite{dallmeier2009generating}. 

Our study focuses on the test-suite based program repair that considers a given test suite as an oracle. 
The test suite should contain at least one initially failing test case that exposes the bug to be repaired
and a number of initially passing test cases that define the expected behavior of the program.
In terms of test-suite based repair, a bug is said to be \emph{fixed} or \emph{repaired} if a repair approach generates
a patch that makes its whole test suite pass. The patch obtained can be referred to as a \emph{test-suite adequate patch} \cite{martinez2016automatic}.

GenProg \cite{weimer2009automatically,le2012genprog,le2012systematic} is one of the most well-known repair approaches for test-suite based program repair. 
This general approach is based on the \emph{redundancy assumption} (i.e., the fix ingredients for a bug already exist
elsewhere in the buggy program); and it uses genetic programming (GP) \cite{koza1992genetic,banzhaf1998genetic} to search for
potential patches that can fulfill the test suite. Although GenProg has been well recognized as a state-of-the-art repair 
approach in the literature, it has aroused certain academic controversies among some researchers. 

First, Qi et al. \cite{qi2014strength} studied to what extent GenProg can benefit from GP. Their
results on GenProg benchmarks \cite{le2012systematic} indicate that just replacing GP in GenProg with random search can improve both 
repair effectiveness and efficiency, thereby questioning the necessity and effectiveness of GP in
automated program repair. Second, an empirical study conducted by Qi et al. \cite{qi2015analysis}
pointed out that the overwhelming majority of patches reported by GenProg are incorrect and 
are equivalent to a single functionality deletion. Here we do not 
focus on the potential incorrectness of the patches that is mainly due to
the weakness of the test suite rather than the repair approaches \cite{monperrus2014critical}. Our 
major concern is that GenProg usually generates nonsensical patches (e.g., a single deletion), which 
challenges the expressive power of GP to produce meaningful or semantically complex repairs. Lastly, a recent large-scale
experiment \cite{martinez2016automatic} showed that an implementation of GenProg for Java (called jGenProg) can find a test-suite
adequate patch for only 27 out of 224 real-world Java bugs, and only five of them were
identified as correct. Obviously, the performance of GenProg for Java is currently far from satisfactory.

Considering these adverse reports about GenProg, it is necessary to revisit the most salient features of
GenProg that qualify it as a well-established repair system. We think there are at least two. 
One is that GenProg can scale to large programs, mainly owing to its patch representation \cite{le2012systematic}. 
Another is that GenProg can potentially address various types of bugs because the expressive power of GP allows for diverse transformations of code.
In particular, GP can change multiple locations of a program simultaneously, so that GenProg is likely to fix 
multi-location bugs that cannot be handled by most of the other repair approaches. 
The scalability of GenProg is visible since it has been widely applied to large real-world software \cite{le2012systematic,martinez2016automatic}, 
making it distinguished from those approaches (e.g., SemFix \cite{nguyen2013semfix} and SearchRepair \cite{ke2015repairing}) that are largely limited to small programs. 
However, as mentioned before, the expressive power of GenProg inherited from GP has not been well supported and validated by recent experimental studies in the literature \cite{qi2014strength, qi2015analysis, martinez2016automatic}. We think it is very important to shed more light on this issue and then address it, which would make
GP really powerful in automated program repair. 

Generally, a successful repair system consists of two key elements \cite{qi2015analysis}: 1) a \emph{search space} 
that contains correct patches; 2) a \emph{search algorithm} that can navigate the search space effectively and efficiently. 
For the search space, GenProg uses the redundancy assumption, which has been largely
validated by two independent empirical studies \cite{martinez2014fix,barr2014plastic}. 
This leaves the search algorithm as a bottleneck that might make GenProg unable to fulfill its potential for generating 
nontrivial patches. The reason could be that the search ability of the underlying GP algorithm in GenProg is not strong enough to really sustain
its expressive power. 

Given this analysis, our primary goal is to improve the effectiveness of the  search via GP for program repair. 
To this end, we present a new GP based repair system for \underline{\textbf{a}}utomated \underline{\textbf{r}}epair of \underline{\textbf{Ja}}va programs, called ARJA. 
ARJA is mainly characterized by a novel patch representation for GP, multi-objective search, a test filtering procedure and several strategies 
to reduce the search space. Our results indicate that an ARJA version that fully follows the redundancy assumption can
generate a test-suite adequate patch for 59 real bugs in four projects of Defects4J \cite{just2014defects4j} as opposed to
only 27 reported by jGenProg \cite{martinez2016automatic}. By manual analysis, we 
find that this ARJA version can synthesize a correct patch for at least 18 bugs in Defects4J as opposed to only 5 by jGenProg. 
To our knowledge, some of the 18 correctly fixed bugs have never been repaired correctly by the other repair approaches.  
Furthermore, ARJA is able to correctly fix several multi-location bugs that are hard to be addressed by 
most of the existing repair approaches. 

The main contributions of this paper are as follows:
\begin{enumerate}
\item The solution representation is a key factor that concerns the performance of GP. 
We propose a novel patch representation for GP based program repair, which properly decouples the search subspaces
of likely-buggy locations, operation types and replacement/insertion code.
\item We propose to formulate automated program repair as a multi-objective optimization problem and employ NSGA-II \cite{deb2002fast} to search 
for potential repairs. 
\item We present a procedure to filter out some test cases that are not influenced by GP from the given test suite, so as to
speed up the fitness evaluation of GP significantly. 
\item We propose to use not only variable scope but also method scope to restrict the number of potential replacement/insertion statements at the destination.
\item We introduce three types of rules which are integrated into three different phases of ARJA search
(i.e., operation initialization, ingredient screening and solution decoding), in order to reduce the search space effectively. 
\item Although our study mainly focuses on improving the search algorithm, we also make an effort to 
enrich the search space reasonably beyond reusing code already extant in the program. To that end, we propose a type matching strategy which can create promising new code for bug fixing
by leveraging syntactic patterns of existing code. 
\item We conduct a large-scale experimental study on 18 seeded bugs and 224 real-world bugs, from which some new findings and insights are 
obtained. 
\item We develop a publicly-available program repair library for Java, which currently includes the implementation of
our proposed approach (i.e., ARJA) and three previous repair 
approaches originally designed for C (i.e., GenProg \cite{le2012systematic}, RSRepair \cite{qi2014strength} and Kali \cite{qi2015analysis}).
It is expected that the library can facilitate further replication and research on automated Java software repair. 
\end{enumerate}

The remainder of this paper is organized as follows. 
In Section \ref{sec-Background and Motivation}, we provide the background knowledge and motivation for our study. 
Section \ref{sec-Approach} describes the proposed repair approach in detail. 
Section \ref{sec-Experimental Design} presents the experimental design
and Section \ref{sec-Experimental Results and Analysis} reports the experimental results obtained. 
Section \ref{sec-Threats to Validity} discusses the threats to validity.
Section \ref{sec-Related Work} lists the related work on test-suite based program repair. 
Finally, Section \ref{sec-Conclusion and Future Work} concludes and outlines directions for future work.

\section{Background and Motivation}
\label{sec-Background and Motivation}
In this section, we first provide background information of ARJA, including
multi-objective genetic programming and the GenProg system. 
Then, we describe the goal and motivation of our study.

\subsection{Multi-Objective Genetic Programming}
\label{sec-Multi-Objective Genetic Programming}

Genetic programming (GP) is a stochastic search technique
which uses an evolutionary algorithm (EA), often derived from a genetic algorithm (GA), to evolve computer programs towards
particular functionality or quality goals. In GP, a computer program (i.e., phenotype) is encoded as a 
genome (i.e., genotype), which can be a syntax tree \cite{koza1992genetic}, an instruction sequence \cite{brameier2007linear}, or other linear and hierarchical data structures \cite{langdon2012genetic};
a fitness function is used to evaluate each genome in terms of how well the corresponding
program works on the predefined task. GP starts with a population of 
genomes that is typically randomly produced and evolves over a series of generations progressively.
In each generation, GP first selects a portion of the current population based on fitness, and then performs
crossover and mutation operators on those selected to generate new genomes which would form the next population. 

Traditionally, the aim of GP is to create a working program \emph{from scratch}, in order to
solve a problem encapsulated by a fitness function. Due to the limited size of successful programs that GP can generate, 
GP research and applications over the past few decades mainly focused on predictive modeling 
(e.g., medical data classification \cite{brameier2001comparison}, energy consumption forecasting \cite{lee2011forecasting} and
scheduling rules design \cite{nguyen2017genetic}), 
where a program is usually just a symbolic expression.
It was not until recently that GP was used to evolve real-world software systems \cite{le2012genprog, le2012systematic,langdon2015optimizing,petke2017specialising}. 
Here, instead of starting from scratch, such GP applications generally take
an \emph{existing} program as a starting point, and then improve it by optimizing its functional properties (e.g., by fixing bugs) \cite{le2012genprog, le2012systematic} or non-functional properties (e.g., execution time and memory consumption) \cite{white2011evolutionary,langdon2015optimizing,petke2017specialising,wu2015deep}. 
This paradigm of applying GP is formally called genetic improvement \cite{petke2017genetic} in the literature.

Moreover, most previous usages of GP only consider a single objective. However, 
there usually exist several competing objectives that need to be optimized simultaneously in a real-world task, 
which can be formulated as a multi-objective optimization problem (MOP). Mathematically,
a general MOP can be stated as 
\begin{equation}\label{eq-MOP}
\begin{aligned}
\min \mathbf{f}(\mathbf{x}) = (f_{1}(\mathbf{x}),  f_{2}(\mathbf{x}), \ldots, f_{m}(\mathbf{x}))^{\mathrm{T}} \\
 \text{subject to  $\mathbf{x} \in \Omega \subseteq \mathbb{R}^{n}$}
\end{aligned}
\end{equation}
$\mathbf{x}$ is a $n$-dimensional decision vector in the decision space $\Omega$, and 
$\mathbf{f}: \Omega \rightarrow \Theta \subseteq \mathbb{R}^{m}$, is an objective vector consisting of $m$ objective functions, which maps the decision space $\Omega$ to
the attainable objective space $\Theta$. 
The objectives in Eq.(\ref{eq-MOP}) are often in conflict with each other (i.e., the decreasing of one objective may lead to the increasing of another), 
so there is typically no single solution that optimizes all objectives simultaneously. To solve a MOP, 
attention is paid to approximating the \emph{Pareto front} (PF) that represents
optimal trade-offs between objectives. The concept of a PF is formally defined as follows.\footnote{In the following we shall assume that the goal is to minimize objectives.}

\begin{definition}[Pareto Dominance]
\label{definition-dominance}
A vector $\mathbf{p}=(p_{1}, p_{2},\\ \ldots, p_{m})^{\mathrm{T}}$ is said to \emph{dominate} another vector $\mathbf{q} = (q_{1}, q_{2},\\ \ldots, q_{m})^{\mathrm{T}}$, denoted by $\mathbf{p} \prec \mathbf{q}$, iff $\forall i \in \{1, 2, \ldots,m\}: p_{i} \leq q_{i}$ and $\exists j \in \{1, 2, \ldots, m\}: p_{j} < q_{j}$.
\end{definition}
\begin{definition}[Pareto Front]
\label{definition-paretofront}
The Pareto front of a MOP is defined as
$ PF := \{\mathbf{f}(\mathbf{x}^{*}) \in \Theta \mid \nexists \mathbf{x} \in \Omega, \mathbf{f}(\mathbf{x}) \prec \mathbf{f}(\mathbf{x}^{*})   \} $.
\end{definition}
From Definition \ref{definition-paretofront}, the PF is a subset of solutions which are 
not dominated by any other solution.

Due to the population-based nature of EAs, they are able to approximate the PF of a MOP in
a single run by obtaining a set of non-dominated objective vectors, from which 
a decision maker can select one or more for their specific needs.
These EAs are called multi-objective EAs (MOEAs). A comprehensive survey of MOEAs can be 
be found in ref. \cite{zhou2011multiobjective}. Considering a suitable multi-objective scenario, multi-objective GP evolves a population of candidate programs for multiple goals using a MOEA approach.

\begin{figure}[htbp]
\centering
\subfloat[Pareto dominance]{\includegraphics[width=0.45\columnwidth]{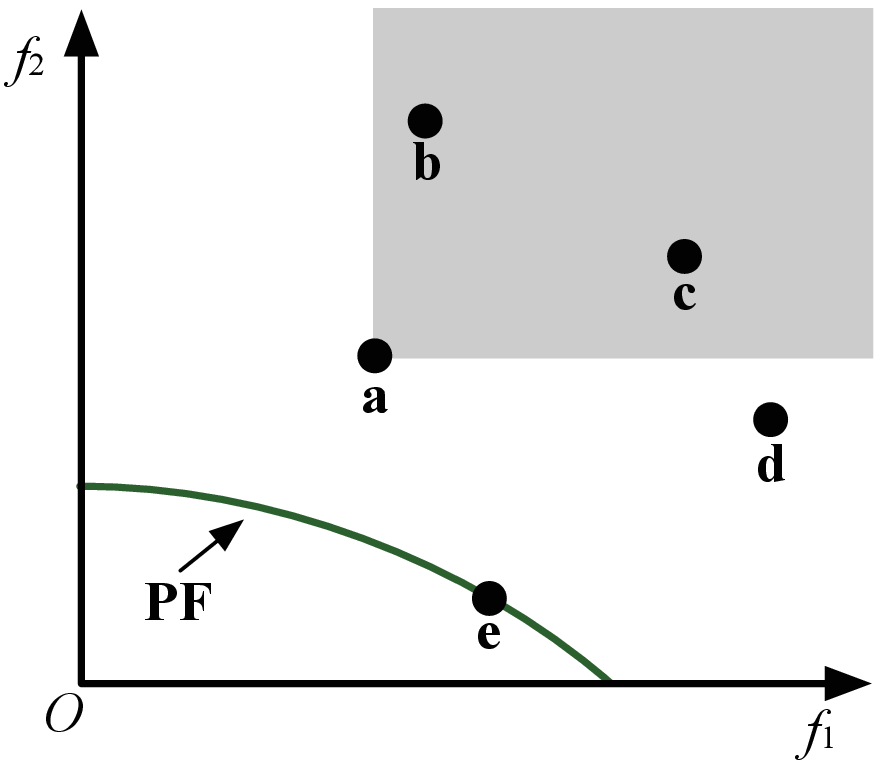}%
\label{fig-dominance}}
\hspace{20pt}
\subfloat[Non-dominated sorting]{\includegraphics[width=0.45\columnwidth]{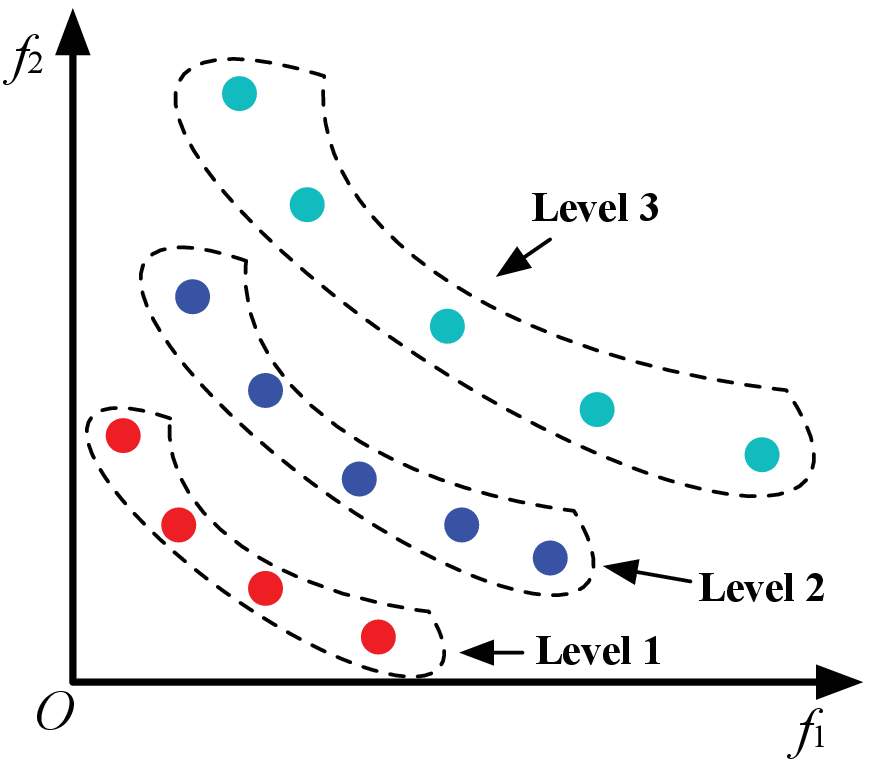}%
\label{fig-sorting}}
\caption{Illustration of Pareto dominance and non-dominated sorting.}
\label{fig-ds}
\end{figure}

Fig. \ref{fig-ds}\subref{fig-dominance} illustrates Pareto dominance for a MOP with two objectives. According to Definition \ref{definition-dominance}, 
all objective vectors within the grey rectangle (e.g., $\mathbf{b}$ and $\mathbf{c}$) are dominated by $\mathbf{a}$, and
$\mathbf{a}$ and $\mathbf{d}$ are non-dominated by each other as $\mathbf{a}$ is better for $f_{1}$ while $\mathbf{d}$ is better for $f_{2}$. 
Because $\mathbf{e}$ is on the PF, no objective vectors in $\Theta$ can dominate it. To provide sufficient selection pressure
toward the PF, many Pareto dominance-based MOEAs, e.g., NSGA-II \cite{deb2002fast}, 
introduce elitism based on non-dominated sorting. Fig. \ref{fig-ds}\subref{fig-sorting} illustrates
the non-dominated sorting procedure, where the union population (combination of current population and offsprings) is
divided into different non-domination levels. The solutions on the first level are obtained by collecting every solution that is not dominated by any other one in the union population. 
To find the solutions on the $j$-th ($j \geq 2$) level, the solutions on the previous $j-1$ levels are first removed, and the same procedure is repeated again. 
The solutions on a lower level will have a higher priority to enter into the next population than those of a higher level.

\subsection{A Brief Introduction to GenProg}
\label{sec-Brief Introduction to GenProg}

\begin{figure}[htbp]
\centering
\subfloat[Patch representation]{\includegraphics[width=0.85\columnwidth]{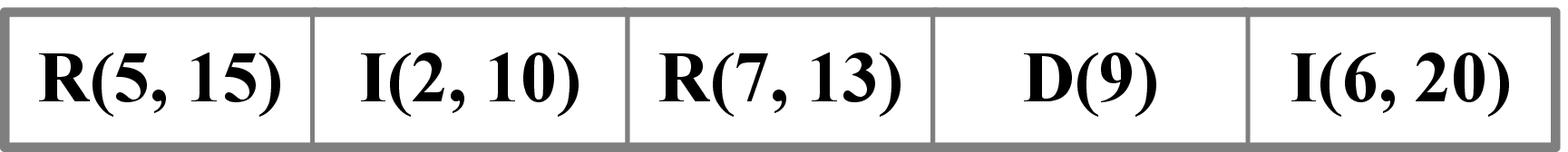}%
\label{fig-genrep}}
\hfill
\subfloat[Crossover]{\includegraphics[width=0.85\columnwidth]{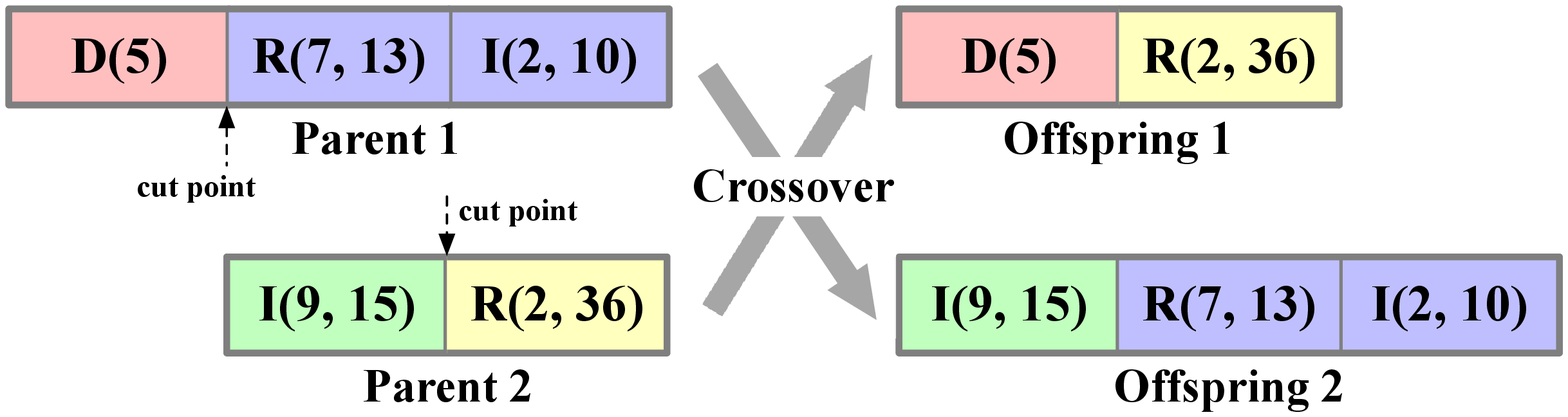}%
\label{fig-genco}}
\hfill
\subfloat[Mutation]{\includegraphics[width=0.85\columnwidth]{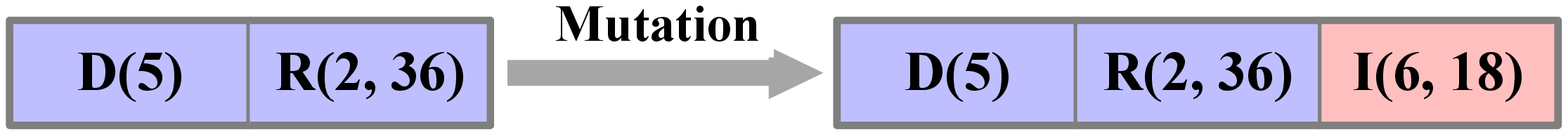}%
\label{fig-genmu}}
\caption{Illustration of patch representation, crossover, and mutation in GenProg. For brevity, ``D'' denotes a delete operation; ``R'' a replace; and ``I'' an insert. The integers denote the AST node numbers
of the corresponding statements.
D(a) means that delete ``a''; R(a, b) means that replace ``a'' with ``b''; I(a,b) means that insert ``b'' before ``a''.}
\label{fig-gcm}
\end{figure}

GenProg \cite{le2012genprog, le2012systematic} is a generic approach that uses GP to automatically find repairs of a buggy program. 
GenProg takes a buggy program as well as a test suite as input and generates one or more test-suite adequate 
patches for output. The test suite is required to contain initially passing tests to model the expected program functionality and at least one
initially failing test to trigger the bug. To obtain a program variant that passes all the given tests, GenProg modifies the buggy program by using a combination of three kinds of 
statement-level edits (i.e., delete a destination statement, replace a destination statement with another statement, and insert another statement before a destination statement).
In the early versions of GenProg \cite{weimer2009automatically, forrest2009genetic, le2012genprog}, 
each genome in the underlying GP is an abstract syntax tree (AST) of the program combined with a weighted path through it. 
However, the AST based representation does not scale to large programs, since the memory consumed by a population of program ASTs is usually unaffordable.
Inspired by ref. \cite{ackling2011evolving}, Le Goues et al. \cite{le2012systematic} addressed the scalability problem of
GenProg by using the \emph{patch representation} instead of the AST representation. 
Specifically, each genome now is represented as a patch, which is stored
as a sequence of edit operations parameterized by AST node numbers (e.g., Replace(7, 13), see Fig. \ref{fig-gcm}\subref{fig-genrep}).
The phenotype of a genome of this representation is 
a modified program obtained by applying the patch to the buggy input program. 

Based on the patch representation, GenProg 
uses either a variant of uniform crossover or single-point crossover to
generate offspring solutions. The experimental results in ref. \cite{le2012representations}
showed that single-point crossover is usually preferable because it can achieve 
a better comprise between success rate and repair effort. The single-point crossover
randomly chooses a cut point each in two parent programs, and the genes beyond the cut points 
are swapped between the two parents to produce two offspring solutions. 
Fig. \ref{fig-gcm}\subref{fig-genco} illustrates the single-point crossover in GenProg. 
In crossover, we can only expect material in the two parents to be differently combined, but not newly generated.

The mutation operator is therefore very important in GenProg, because it is responsible 
for introducing new edit operations into the population. To conduct the mutation on a solution, 
first each potential faulty statement is chosen as a destination statement with a probability of
mutation rate weighted by its suspiciousness. Once
a destination statement is determined, an operation type is randomly chosen
from three types (i.e., ``Delete'', ``Replace'' and ``Insert''). In case of 
``Replace'' or ``Insert'', a second statement (i.e., replacement/insertion code)
is randomly chosen from those statements which only reference variables in the \emph{variable scope} at the destination
and are visited by at least one test. Each edit operation created in this way is appended to
a list of edits in the solution under mutation. Fig. \ref{fig-gcm}\subref{fig-genmu} illustrates the mutation operator in GenProg.

The overall procedure of GenProg is summarized as follows. First, by running all tests, GenProg localizes 
potential buggy statements and gives each of them a weight measuring its suspiciousness.
Then, GP searches to  obtain an initial population by independently mutating $N$ (the population size) copies of the empty patch. 
In each generation of GP, GenProg uses tournament selection based on fitness (i.e., the ability to pass the tests) to select 
$N / 2$ parent solutions for mating from the current population, and conducts crossover (as in Fig. \ref{fig-gcm}\subref{fig-genco}) on the
parents to generate $N/2$ offspring solutions. 
Afterwards, it conducts one mutation (as in Fig. \ref{fig-gcm}\subref{fig-genmu}) on each 
parent and each offspring. The parents together with the offsprings will form the
next population. The GP loop is terminated when a program variant passes all given tests or
another termination criterion is reached.

\subsection{Goal and Motivation}
\label{sec-Goal and Motivation}

Our overall goal in this study is to develop a more powerful GP based system for automated repair of Java programs. To this end, 
we conduct an analysis of the potential limitations of GenProg so as to guide the design of our new system. 
There are several deficiencies in GenProg that motivated us to pursue this goal, which are discussed as follows. 

\subsubsection{High-Granularity Patch Representation}
\label{sec-High-Granularity Patch Representation}

In GenProg, each gene in the patch representation (see Fig. \ref{fig-gcm}\subref{fig-genrep})
is a high-granularity edit operation in which the operation type, likely-buggy location (i.e., destination statement), and 
replacement/insertion code are \emph{invisible} to the crossover (see Fig. \ref{fig-gcm}\subref{fig-genco}) and mutation (see Fig. \ref{fig-gcm}\subref{fig-genmu}) operators. 
Manipulating such high-level units via GP would hinder the efficient recombination of genetic information between solutions. 
This is mainly because  good partial information of an edit operation (e.g., a promising operation type, an 
accurate faulty location, and a useful replacement/insertion code)
cannot be propagated from one solution to others.  
For illustration purposes, suppose there is a bug that requires two
edit operations to be repaired: D(5), R(2, 10), 
and there are two candidate solutions in the population that are
the same as ``Parent 1'' and ``Parent 2'' in Fig. \ref{fig-gcm}\subref{fig-genco} respectively. 
As can be seen, the two candidate solutions together contain
all the material to compose the correct patch. The crossover in GenProg
can easily propagate 
the desired edit D(5) in ``Parent 1'' to offspring solutions. However, because
such crossover does not introduce any new edit, it cannot produce R(2,10), even though 
I(2, 10) in ``Parent 1'' and R(2, 36) in ``Parent 2'' are both one modification away and
their desired partial information can be obtained from each other. The mutation in GenProg 
creates new edits from scratch, where the operation types and 
replacement/insertion code are just randomly chosen from all those available. Thus, there is only a
slim chance for mutation to produce exactly R(2,10).

\begin{figure}[htbp]
\centering
\subfloat[Patch representation]{\includegraphics[width=0.85\columnwidth]{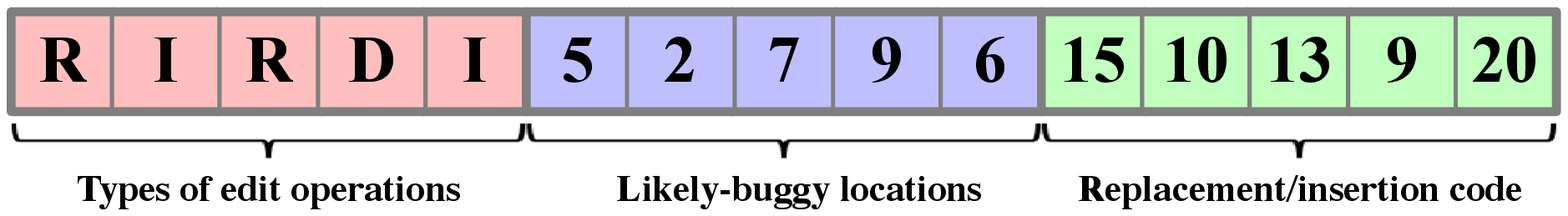}%
\label{fig-newrep}}
\hfill
\subfloat[Crossover]{\includegraphics[width=0.85\columnwidth]{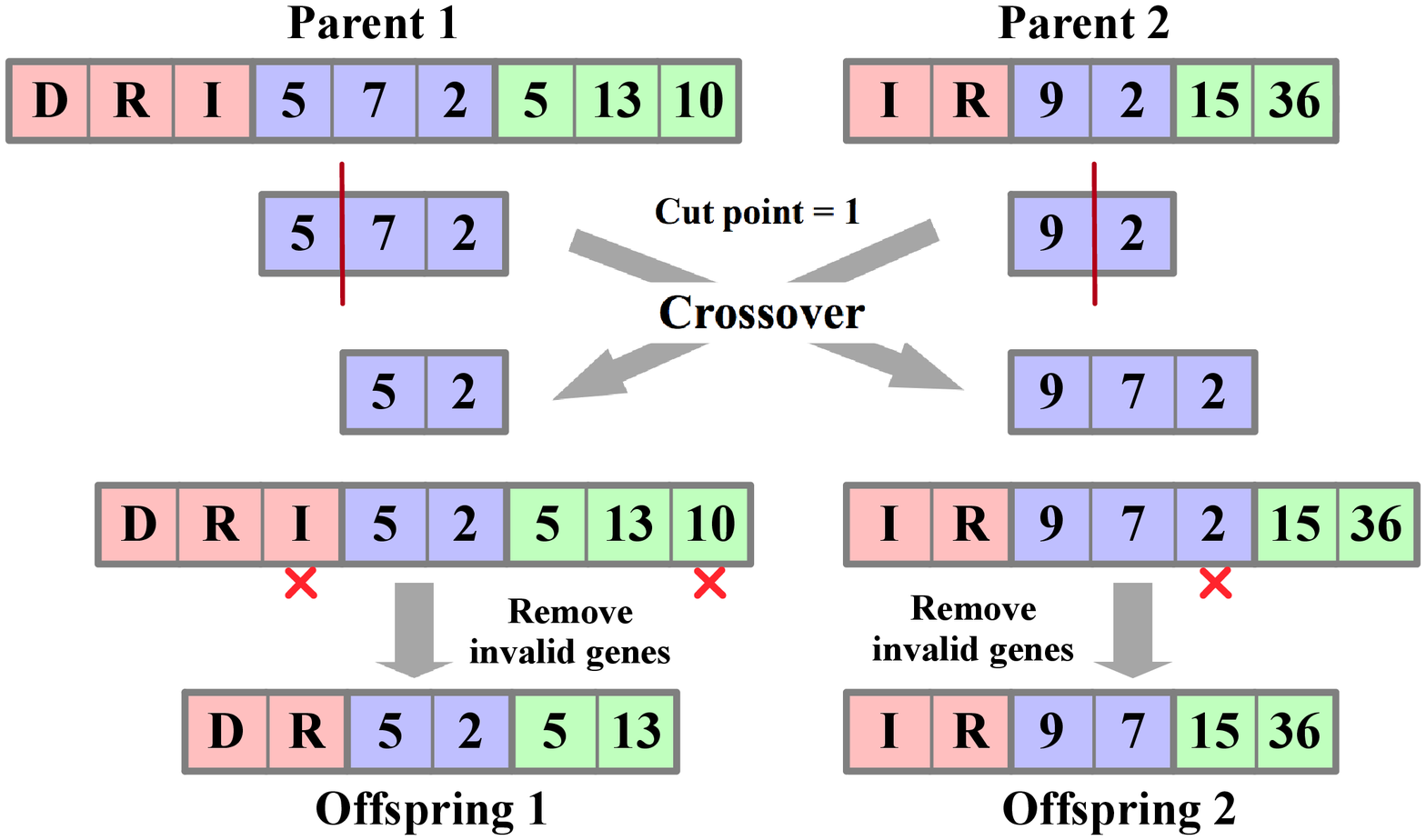}%
\label{fig-newco}}
\caption{Illustration of patch representation and crossover introduced by Oliveira et al. \cite{oliveira2016improved}.} \label{fig-newrep}
\label{fig-newrc}
\end{figure}

Recently, Oliveira et al. \cite{oliveira2016improved} noted the above limitation of GenProg
and presented a lower-granularity patch representation, which decouples the
three subspaces corresponding to three kinds of partial information in an edit operation. 
Using this representation, the patch represented in Fig. \ref{fig-gcm}\subref{fig-genrep} can
be reformulated as that shown in Fig. \ref{fig-newrc}\subref{fig-newrep}, where the representation is divided into three different 
parts: the first part is a list of operation types, the second a list of likely-buggy locations, and the third a list
of replacement/insertion code. Such a representation makes the search explicitly explore
three subspaces and thus overcomes GenProg's drawback of over-constraining the search ability to some extent, 
but it can also lead to new problems.

One problem is that  crossover  becomes less elegant. 
Fig. \ref{fig-newrc}\subref{fig-newco} illustrates a kind of crossover suggested in ref. \cite{oliveira2016improved}. 
The crossover first randomly chooses one part in the representation, and conducts single-point crossover only on
that part  keeping the other two parts unchanged. However, due to different numbers of genes in the three parts after crossover,
there exist some invalid genes that should be removed to obtain a final offspring solutions. The removal
of invalid genes will potentially result in information loss. To relieve this issue, 
Oliveira et al. \cite{oliveira2016improved} introduced a memorization scheme to reuse  invalid genes. However, this
additional procedure will inevitably complicate the crossover and increase the computational burden. 

Another problem is more essential. That is,  crossover on
such a representation  exchanges  information of operation types and
replacement/insertion code between different likely-buggy locations very frequently. 
However, this situation is indeed not desirable, because every likely-buggy location has its own
syntactic/semantic context and preferable operation types and replacement/insertion code can vary a lot. 
Moreover, due to scoping issues, just the available replacement/insertion code can be quite different 
at different likely-buggy locations, so exchanging replacement/insertion code between such locations
usually even results in an uncompilable program variant. For example, in Fig. \ref{fig-newrc}\subref{fig-newco},
the statement numbered 13 is originally the replacement code at the location 7, whereas after crossover, it
becomes that at the location 2; however, certain variables in this statement can be out of the scope at the new location.  
 
Our study aims to propose a novel lower-granularity patch representation that can address
the limitations of GenProg's representation while avoiding the above two problems caused by the representation introduced in ref. \cite{oliveira2016improved}.

\subsubsection{High Complexity of Repairs}
\label{sec-High Complexity of Repairs}

GP based repair methods (e.g., GenProg) can carry out complex transformations of original code via GP, so they
have the potential to repair a buggy program that really requires many modifications. However, the great expressive power of GP
can also pose the risk of finding a repair that is much more complex than necessary. 
In practice, a simpler patch is usually preferred, for two reasons: (1) 
According to Occam's razor, a simpler patch could have better 
generalization ability to unseen test cases, and (2) 
such a patch is generally more comprehensible and maintainable for humans.  
Hence, it is beneficial to direct a program repair tool to generate simple or small patches. 
A recent effort toward this goal can be seen in ref. \cite{mechtaev2015directfix}, where a
semantics-based repair method was proposed to find simple program repairs. 

Given that  existing GP based repair methods (e.g., GenProg) do not explicitly take into account the simplicity of a patch 
during search,
our study aims to incorporate a consideration for the simplicity of repairs into the search process of GP, so as to 
introduce search bias towards small patches.

\subsubsection{Expensive Fitness Evaluation}
\label{sec-Expensive Fitness Evaluation}

In GenProg, all given tests need to be run in order to evaluate the fitness of 
a solution accurately. However it is usually computationally expensive to run all the associated tests of 
real-world software. For example, there are more than 5200 JUnit tests in the current Commons Math project,\footnote{Apache Commons Math, http://commons.apache.org/math} and 
it would take about 200 seconds to execute all of them for just one fitness evaluation
on an Intel Core i5 2.9GHz processor with 16Gb of memory. 
Expensive fitness evaluations will limit the use of a reasonably large number of generations or population size in GP, 
thereby greatly limiting the potential of GP for program repair. 
To relieve this problem, Fast et al. \cite{fast2010designing} proposed to just use a random 
sample of given tests for each fitness evaluation. Although that strategy 
can increase efficiency, it will unavoidably reduce the precision of the search. 

Our study argues that not all  given tests are necessary for fitness evaluation. In fact, 
many can be omitted in order to speed up the fitness evaluation, but without affecting the precision of search.

\subsubsection{Incomplete Scope}
\label{sec-Incomplete Scope}
To follow the principle of high cohesion, modern Java software systems 
generally involve a considerable number of method invocations (e.g.,  up to 20,000 in the Commons Math project). 
However, when choosing replacement/insertion code, GenProg checks the validity of the code only 
according to the variable scope at destination, without considering the method scope. 
This practice may be effective for  C programs considered in ref. \cite{le2012systematic}, but is not
good enough for  real-world Java programs investigated in our study. 
For example, for a buggy version of  the Commons Math project, we find that,
among the available replacement/insertion statements (satisfying the variable scope only) at each
destination, on average about 40\%  invoke invisible methods and are indeed invalid. 

Hence, our study proposes to consider both  variable scope
and method scope when choosing valid replacement/insertion code for 
a likely-buggy location, so as to improve the success rate of compiling 
the modified programs. 

\subsubsection{Limited Utilization of Existing Code}
\label{sec-Limited Utilization of Existing Code}
Today, large Java projects are commonly developed by
many programmers, each of whom is responsible for only one
or several modules. Although the names of important APIs or even
field variables can be determined in the software design phase, 
the names of local variables and private methods are generally chosen based on the
preference of the responsible programmer, which leads to the fact that even variables or methods with similar functions
can have different names in different Java files. 
Thus, for a likely-buggy location, it is sometimes possible that we can make an invalid statement  
become its hopeful replacement/insertion code by replacing the
invisible variables or methods with similar ones in the scope. In other words,
the underlying pattern in a statement other than the statement itself can also be exploited to acquire useful 
replacement/insertion code. 
GenProg does not create any new code, in which the replacement/insertion code
is just taken from somewhere else in the buggy program without change. 
This practice may fail to make the most of the existing code.

Our study aims to present a strategy that can exploit the pattern of 
the existing code appropriately, so as to create some new replacement/insertion statements
that are potentially useful.

\subsubsection{Lack of Adequate Prior Knowledge}
\label{sec-Lack of Adequate Prior Knowledge}

GenProg can conduct any deletion, replacement or insertion operations 
on the possibly faulty statements, provided that the replacement/insertion 
code meets the variable scope. However, from the view of an experienced programmer, 
some operations indeed make little sense, which is mainly due to the following two reasons.

One reason is that although a replacement/insertion statement conforms to
the scope of variables and methods at a destination, it can still violate other Java 
language specifications when it is pasted to that place. Another reason is that certain operations
either disrupt the  program considered too much or have no effect at all. 
For example, in Fig. \ref{fig-vd}, if we delete the variable declaration statement (at line 1092),  
all the remaining statements will be invalidated immediately since they all reference the variable 
\lstinline[language=Java]!cloned!. Moreover, even if a variable declaration statement should be deleted, leaving it as a redundant statement generally does not influence the correctness of the program.  
Thus the deletion operation here is not desired and should be disabled. 


\begin{figure}[htbp]
\centering
\begin{lstlisting}[firstnumber=1092]
final StrTokenizer cloned = (StrTokenizer) super.clone();
if (cloned.chars != null) {
	cloned.chars = cloned.chars.clone();
}
cloned.reset();
return cloned;
\end{lstlisting} 
\caption{The code snippet excerpted from the Commons Lang project.\protect\footnotemark}
\label{fig-vd}
\end{figure}
\footnotetext{Apache Commons Lang, http://commons.apache.org/lang}

Our study aims to encode such prior knowledge of programmers as
a number of rules, which can be integrated into the proposed repair method flexibly.
We expect that the search space of GP can be reduced effectively with these rules. 
Note that our aim is very different from that of ref. \cite{tan2016anti}.
The rules considered in our study disallow definitely unnecessary operations rather than 
likely unpromising ones, so they generally do not restrict the expressive power of GP.

\section{Approach}
\label{sec-Approach}

This section presents our generic  approach to automatically finding the test-suite adequate patches via multi-objective GP.
This approach is implemented as a tool called ARJA that repairs Java code.

\subsection{Overview}
\label{sec-Overview}

\begin{figure*}[htbp]
\centering
        \includegraphics[scale=0.43]{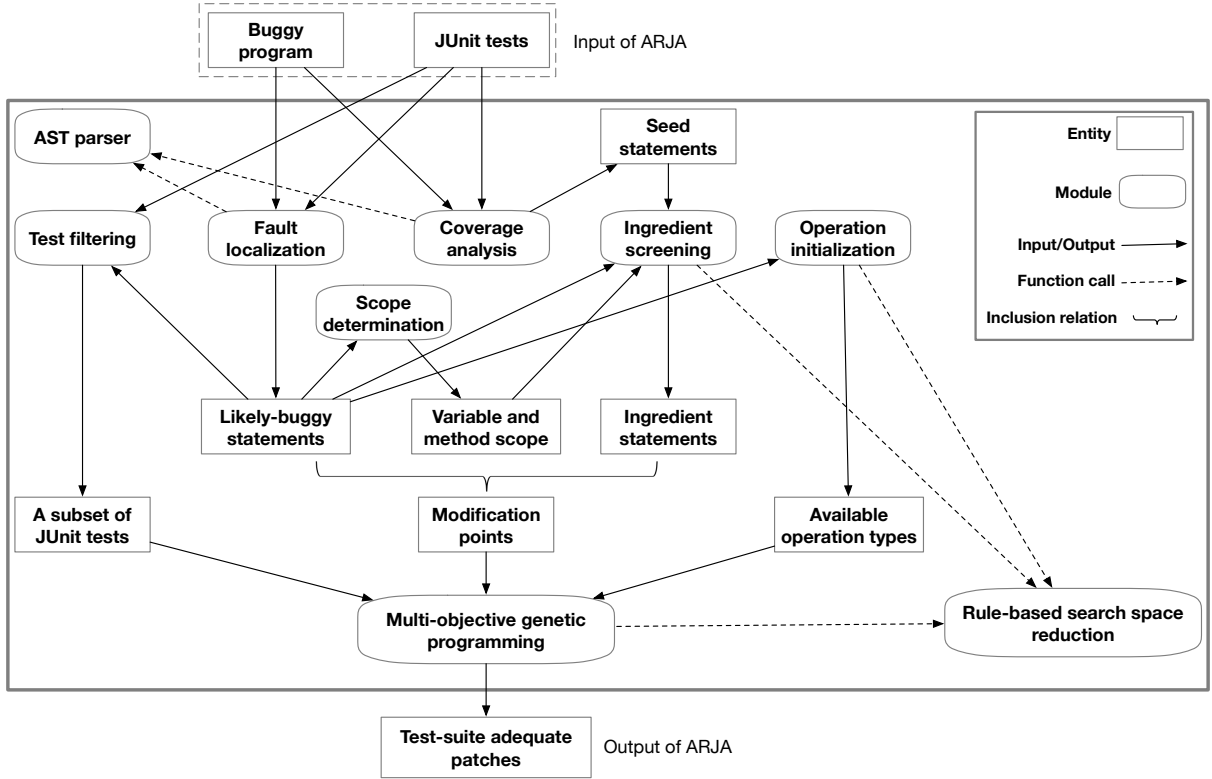}
\caption{Overview of the proposed automated program repair approach,  i.e., ARJA.} \label{fig-overview}
\end{figure*}

In a nutshell, ARJA works as depicted in Fig.  \ref{fig-overview}. 
ARJA takes a buggy program and a set of associated JUnit tests as the input. 
Among the tests, at least one \emph{negative} (i.e., initially failing) test is required to be included, which exposes
the bug to be fixed. All the remaining are \emph{positive} (i.e., initially passing) tests, which describe
the expected program behavior.  The basic goal of ARJA is to modify the program so that all 
tests pass. Its process is composed of the following main steps.  

Given the input, a fault localization technique is used to 
identify  potentially buggy statements which are to be manipulated
by GP. Meanwhile,  coverage analysis is conducted to record every statement that
is visited by any JUnit test. These statements collected
in the coverage analysis (referred to as \emph{seed statements} in ARJA)
provide the source of the replacement/insertion code (referred to as
\emph{ingredient statements} in ARJA).
Note that fault localization and coverage analysis both require the Eclipse AST parser to transform the 
line information of  code to the corresponding Java statements.

Once the likely-buggy statements are identified, they will be put to use
immediately in two ways.
(1) positive tests unrelated to these statements are filtered
out from the original JUnit test suite, so that a reduced set of tests
can be obtained for further use. 
(2) the scope of variables and methods
is determined for the location of each of these statements. 

Then the ingredient statements for each  likely-buggy statement considered are
selected from the seed statements in view of the current variable and method scope. 
For convenience, a likely-buggy statement along with the scope at its location and its corresponding ingredient 
statements is called a \emph{modification point} in short.

Before entering into the genetic search, the types of 
operations on  potentially buggy statements should be defined in advance. 
Similar to GenProg \cite{le2012systematic}, ARJA also uses three kinds of operations: \emph{delete}, \emph{replace}, and 
\emph{insert}.  More specifically, for each likely-buggy statement, ARJA either deletes 
it, replaces it with one of its ingredient statements, or inserts one of its ingredient statements before it.
Note that although only three operation types are currently used,  users can add 
other possible types \cite{martinez2015mining} into ARJA easily due to its flexible design. 
 
With a number of modification points, a subset of  original JUnit tests, and the available 
operation types in place, ARJA encodes a program patch with a novel GP representation. 
Based on this new representation, a MOEA is 
employed to evolve the patches by simultaneously minimizing the failure rate on
tests and  patch size. Finally, the non-dominated solutions obtained with 
a failure rate of 0 are output as  test-suite adequate patches.

Notably, ARJA is also characterized by a module that reduces
the search space based on some specific rules. These rules 
can be divided into three different types which are specially designed
for operation initialization, ingredient screening and decoding in multi-objective GP,
respectively. Applying such rules allows the modified program to be compiled successfully
with a higher probability, while focusing the search  on more promising regions of the search space.

\subsection{Fault Localization and Coverage Analysis}
\label{sec-Fault Localization and Coverage Analysis}

For fault localization, ARJA uses an existing spectrum-based technique called Ochiai \cite{abreu2006evaluation}.
It computes a suspiciousness measure of a line of code ($lc$) as follows:
\begin{equation}\label{eq-ochiai}
\begin{aligned}
susp(lc) = \frac{N_{CF}}{\sqrt{N_{F} \times (N_{CF} + N_{CS})}}
\end{aligned}
\end{equation}
where $N_{CF}$ and $N_{CS}$ are the number of negative tests and
 positive tests that visit the code $lc$, respectively, and  
$N_{F}$ are the total number of negative tests. Fault
localization analysis returns a number of potentially faulty lines,  each represented as a tuple
$(cls, lid, susp)$.  $cls$ and $lid$ are the name of the Java class and the line number 
in this class, respectively, which are used to identify a line uniquely, and $susp \in [0,1]$
is the corresponding suspiciousness score.

To look for seed statements, ARJA implements a strategy presented in ref. \cite{le2012systematic} to 
reduce the number of seed statements and to choose those more related to
the given JUnit tests. That is,  coverage analysis
is conducted to find the lines of code that are visited by
at least one test, each of which forms a tuple $(cls, lid)$. 

After the above phases, the Eclipse AST parser is used to parse the potentially faulty lines and
seed lines into the likely-buggy statements and seed statements respectively.
For duplicate seed statements, only one of them is recorded. 

Note that in ARJA, we do not consider all  potentially faulty statements. Instead, only a part of them
are selected according to their suspiciousness in order to reduce the search space. 
The number of statements can be controlled in ARJA by either of  two parameters denoted 
$\gamma_{\min}$ and $n_{\max}$. $\gamma_{\min}$ 
quantifies the minimum suspiciousness score for statements to be considered, while $n_{\max}$ determines that at most $n_{\max}$
likely-buggy statements with highest suspiciousness are chosen. If both $\gamma_{\min}$ and $n_{\max}$ are set,
ARJA uses the smaller number determined by either of them.

\subsection{Test Filtering}
\label{sec-Test Filtering}
In ARJA, we propose to ignore those positive tests that are
unrelated to the considered likely-buggy statements. 
To be specific, for each positive test, we record all the lines of code 
covered during its execution, and if these lines do not
include any of the lines associated with the likely-buggy statements selected, we can safely filter
out this positive test. This strategy usually 
enables us to obtain a much smaller test suite compared to the original one,
which can significantly speed up the fitness evaluation in GP search without
influencing its effectiveness. 

\begin{figure}[htbp]
\centering
        \includegraphics[scale=0.43]{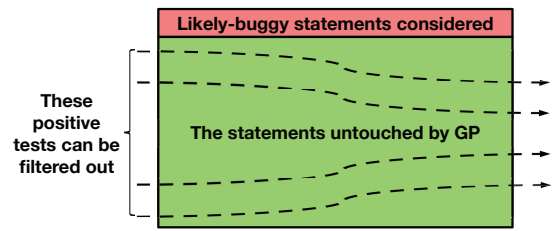}
\caption{Illustration of the execution path of the positive test that can be filtered out.} \label{fig-testfilter}
\end{figure}

The rationale for test filtering can be explained as follows:
since  positive tests ignored by this procedure do not touch
any code used by GP, every program variant 
produced by GP will always pass these test cases once it can 
be compiled and run successfully. 
Thus, after discarding such positive tests, the relative superiority or inferiority in fitness between any two GP individuals will be kept the same, and the search behavior of GP will
not be changed. Fig. \ref{fig-testfilter} illustrates the principle of test filtering.

\subsection{Scope Determination}
\label{sec-Scope Determination}

For each likely-buggy statement considered, most  seed statements cannot 
become an ingredient statement. This is mainly because these seed statements access  variables or methods that are invisible at the location of the likely-buggy statement. 
To identify as many of them as possible, we have to determine the scope (i.e., all the visible variables and methods) at that location.

Note that unlike GenProg \cite{le2012systematic}, ARJA  considers not only the \emph{variable} scope but also the \emph{method} scope, which
can improve the chance of the modified Java program to being compiled successfully.
The necessity of this practice for Java has been discussed in Section \ref{sec-Incomplete Scope}.

Suppose $Cls$ and $Med$ are the class and the method where a likely-buggy statement appears, respectively.
According to the Java language specification, ARJA collects three kinds of variables to constitute the variable scope: the visible field variables in $Med$, the parameter variables of $Med$, 
and the local variables in $Med$ defined before the location of the likely-buggy statement. Among them, 
the first kind of variables has three sources: the field variables declared in $Cls$, the field variables inherited from the parent classes of $Cls$, and the field variables
declared in the outer classes (if they exist) of $Cls$. 
As for the method scope, ARJA collects the visible methods in $Med$, which have three sources similar to the visible field variables. 

Note that besides the variable and method names, ARJA also records their corresponding type information and modifiers to make the scope more accurate. 
For a method, the type information includes both  parameter types and the return type.

\subsection{Ingredient Screening}
\label{sec-Ingredient Screening}

This procedure aims to select the ingredient statements for each likely-buggy statement considered. 
In this phase, ARJA first adopts the \emph{location awareness} strategy introduced in ref. \cite{martinez2016astor}. 
This strategy defines three alternative ingredient modes (i.e., \emph{File}, \emph{Package}, \emph{Application}), which are used
to specify the places where ingredients are taken from. Suppose a likely-buggy statement is located in the file $Fl$ that belongs
to the package $Pk$, then the ``File'' and ``Package'' modes mean that this likely-buggy statement can only take
its ingredient statements from $Fl$ and $Pk$, respectively.
The ``Application'' mode means that the ingredient statements can come from anywhere in the entire buggy program. 
Compared to the ``Application'' mode, the other two modes can significantly restrict the space of ingredients, which may 
help to find the repairs faster or find more of them.  

With the location awareness strategy incorporated, ARJA provides the following two alternative approaches for ingredient screening, namely a direct approach and a type matching based approach. 

\subsubsection{Direct Approach}
\label{sec-Direct Method}

The direct approach works as follows. For each considered likely-buggy statement,
all the seed statements are examined one by one. 
If a seed statement does not come from the place specified by the ingredient mode, 
it will just be ignored.  Otherwise, we extract the variables and methods accessed by this seed statement. 
For example, for the following statement: 
\begin{lstlisting}[numbers=none,frame =none]
ret = 1.0 - getDistribution().beta + b.regularizedBeta(getProbability(), this.x + 1.0, getTrials() - this.x);
\end{lstlisting} 
the extracted variables include \lstinline[language=Java]!ret!, \lstinline[language=Java]!b!, and \lstinline[language=Java]!x!;
and the extracted methods are \lstinline[language=Java]!getDistribution!, \lstinline[language=Java]!getProbability! and \lstinline[language=Java]!getTrials!.  
Note that it is not necessary to consider the variable \lstinline[language=Java]!beta! and the method \lstinline[language=Java]!regularizedBeta!, because
their accessibility generally only depends on the visibility of \lstinline[language=Java]! getDistribution! and \lstinline[language=Java]!b!, respectively. 

For each extracted variable/method, we check whether the one with the same name and the compatible type exists in the variable/method scope (determined in Section \ref{sec-Scope Determination}). 
Only when all of them have the corresponding ones in the variable/method scope, this seed statement can become an ingredient statement of the likely-buggy statement.

\subsubsection{Type Matching Based Approach}
\label{sec-Type Matching Based Method}

As mentioned in Section \ref{sec-Limited Utilization of Existing Code}, it could be demanding for a seed statement to 
only access the variables/methods visible at the location of the likely-buggy statement. Indeed, the pattern of a seed statement can 
sometimes also be useful. 

To exploit such patterns, the type matching based approach goes a step further compared to the direct approach. When certain variables or methods extracted from a seed statement cannot be found
in the variable or method scope, the type matching based approach does not discard this seed statement immediately. Instead, 
it tries to map each variable or method out of  scope to one with the compatible type in  scope.
To restrict the complexity, we follow two guidelines in type matching: 1)
Different variables/methods in a seed statement must correspond to different ones in the scope; 2) If there is more than one variable/method
with a compatible type, the one with the same type is preferred.

\begin{figure}[htbp]
\centering
        \includegraphics[scale=0.53]{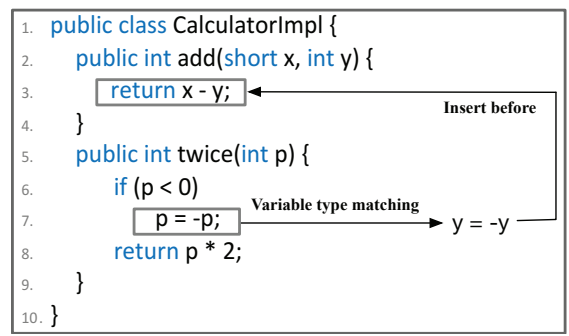}
\caption{Illustration of the type matching for variables.} \label{fig-vmat}
\end{figure}

\begin{figure}[htbp]
\centering
        \includegraphics[scale=0.53]{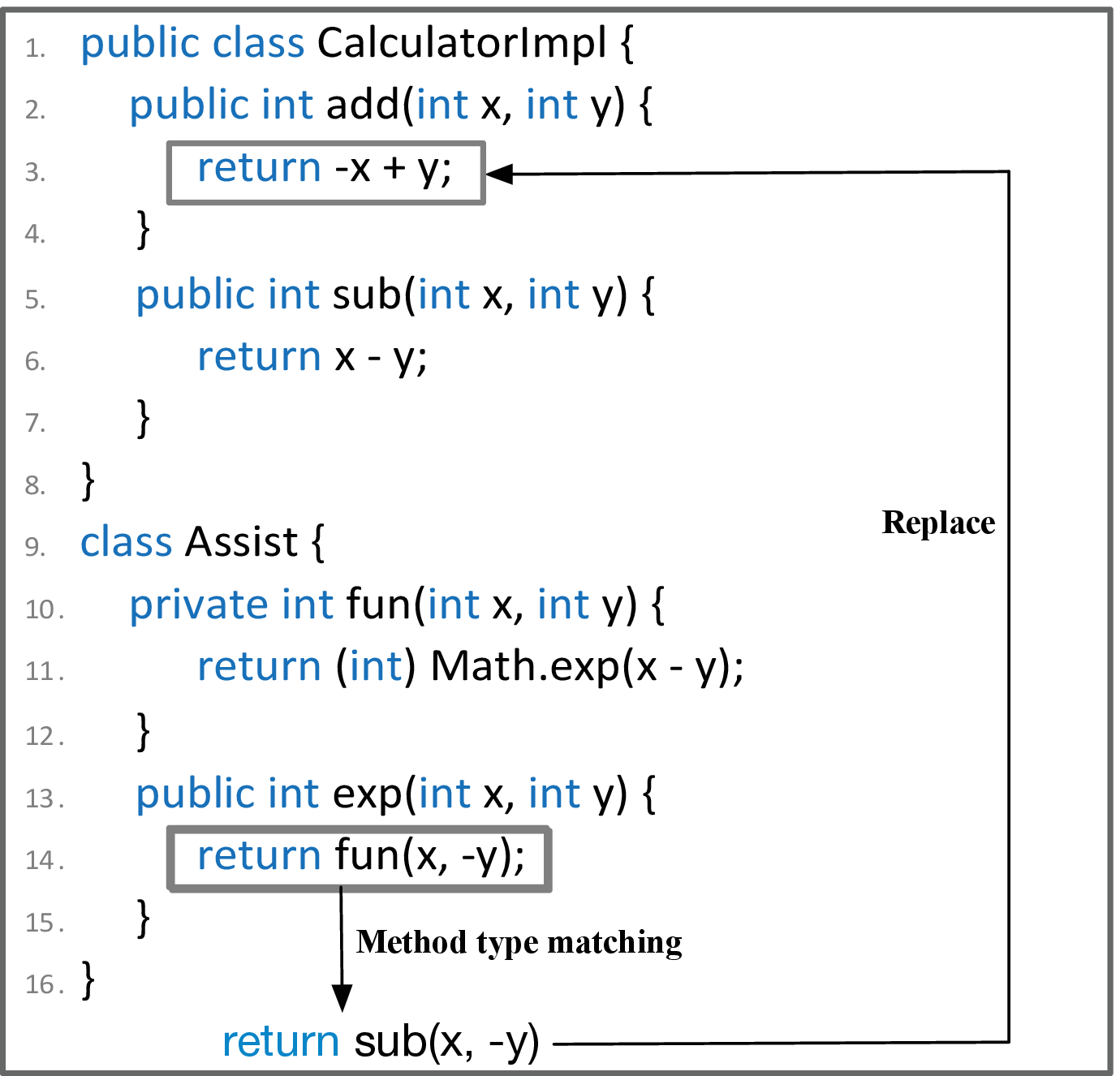}
\caption{Illustration of the type matching for methods.} \label{fig-mmat}
\end{figure}

If the type matching is successful, the modified seed statement will become an ingredient statement. 
Fig. \ref{fig-vmat} and Fig. \ref{fig-mmat} illustrate how type matching works for variables and methods respectively, using toy programs.


In Fig. \ref{fig-vmat}, the statement at line 3 is a faulty one, and the bug can be
fixed by inserting \lstinline[language=Java]!y=-y! before this statement. 
ARJA cannot repair this fault without type matching since no such fix ingredient exists in the current program.
However, if type matching is enabled, \lstinline[language=Java]!y=-y! can be generated via
a seed statement \lstinline[language=Java]!p=-p!, by mapping the variable 
\lstinline[language=Java]!p! to \lstinline[language=Java]!y!.
Similarly, in Fig. \ref{fig-mmat}, the method \lstinline[language=Java]!fun! is mapped to \lstinline[language=Java]!sub!, and an
ingredient statement \lstinline[language=Java]!sub(x,-y)! is generated via \lstinline[language=Java]!fun(x,-y)!, which can be used to fix the bug at line 3.

\subsection{Evolving Program Patches}
\label{sec-Evolving Program Patches}

Suppose that we have selected $n$ likely-buggy statements using the 
procedure in Section \ref{sec-Fault Localization and Coverage Analysis}, 
and each of them has a set of ingredient statements found by the 
method in Section \ref{sec-Ingredient Screening}. Thus, we have $n$
modification points, each of which has a likely-buggy statement and its corresponding ingredient statements.

With the $n$ modification points along with the available operation types
and a reduced set of JUnit tests (obtained in Section \ref{sec-Test Filtering}), 
we can now encode a program patch as a genome and evolve a population of such tentative
patches via multi-objective GP.
The details are given as follows.

\subsubsection{Solution Representation}
\label{sec-Solution Representation}
To encode a patch, we first order the $n$ modification points in some manner. For
the $j$-th modification point, where $j=1,2,\ldots,n$, the corresponding set of 
ingredient statements is denoted by $I_{j}$ and the statements in $I_{j}$ are also
ordered arbitrarily. Moreover, the set of operation types is denoted by $O$ and the elements
in $O$ are numbered starting from 1. Note that the ID number for each modification point, each statement in $I_{j}$,
or each operation type in $O$ is fixed throughout the evolutionary process.

In ARJA, we propose a new patch representation that can decouple the search subspaces of likely-buggy locations, 
operation types and ingredients perfectly. Specifically, each solution
can be represented as $\mathbf{x}=(\mathbf{b},\mathbf{u},\mathbf{v})$, which 
contains three different parts and each part is a vector with size $n$. 

\begin{figure}[htbp]
\centering
        \includegraphics[scale=0.45]{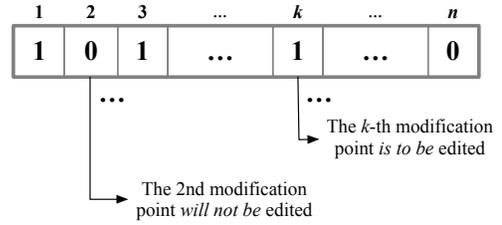}
\caption{Illustration of the first part (i.e., $\mathbf{b}$) of the representation.} \label{fig-bv}
\end{figure}

The first part, denoted by $\mathbf{b}=(b_{1}, b_{2}, \ldots, b_{n})$, is a binary vector with $n$  bits $b_{j}$ ($b_{j}\in\{0, 1\}, j=1,2,\ldots,n$). 
$b_{j}$ indicates whether or not the patch $\mathbf{x}$ chooses to edit the likely-buggy statement in the $j$-th modification point.
Fig. \ref{fig-bv} illustrates the representation of $\mathbf{b}$.

\begin{figure}[htbp]
\centering
        \includegraphics[scale=0.40]{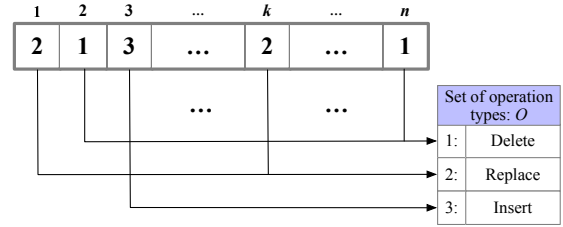}
\caption{Illustration of the second part (i.e., $\mathbf{u}$) of the representation.} \label{fig-uv}
\end{figure}

The second part, denoted by $\mathbf{u}=(u_{1}, u_{2}, \ldots, u_{n})$, is a vector with $n$ integers, where $u_{j} \in [1, |O|], j = 1, 2, ,\ldots, n$.
$u_{j}$ means that the patch $\mathbf{x}$ chooses the $u_{j}$-th operation type in the set $O$ for the $j$-th modification point. 
In Fig. \ref{fig-uv}, we illustrate the representation of $\mathbf{u}$. 

\begin{figure}[htbp]
\centering
        \includegraphics[scale=0.35]{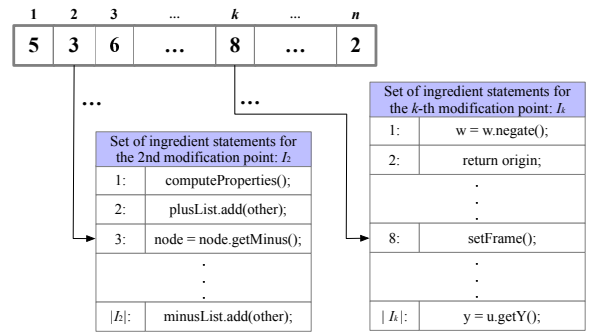}
\caption{Illustration of the third part (i.e., $\mathbf{v}$) of the representation.} \label{fig-vv}
\end{figure}

Similar to $\mathbf{u}$, the third part (i.e., $\mathbf{v}=(v_{1}, v_{2}, \ldots, v_{n})$) is also a vector with $n$ integers,
where $v_{j} \in [1,|I_{j}|], j = 1, 2, \ldots, n$. $v_{j}$ indicates that the patch $\mathbf{x}$ chooses the $v_{j}$-th ingredient statement in
the set $I_{j}$ for the $j$-th modification point. Fig. \ref{fig-vv} illustrates the representation of $\mathbf{v}$.

As can be seen, $b_{j}$, $u_{j}$ and $v_{j}$ together determine what the patch $\mathbf{x}$ does
to the $j$-th modification point. For example, for the patch represented in Figs. \ref{fig-bv}, \ref{fig-uv} and \ref{fig-vv}, 
it replaces the likely-buggy statement in the $k$-th modification point with \lstinline[language=Java]!setFrame()!.
Suppose the operation types in $O$ are numbered as in Fig. \ref{fig-uv}, 
the whole procedure to apply a patch $\mathbf{x}$ (i.e., the decoding procedure)
is described in Algorithm \ref{alg-patch}.
\begin{algorithm}[h]
\label{alg-patch}
\small
\LinesNumbered 
 \KwIn{$n$ modification points; the set of operation types $O$;\\ a patch $\mathbf{x}=(\mathbf{b},\mathbf{u},\mathbf{v})$.}
 \KwOut{A modified program.}
\For {$j=1$ \KwTo $n$} {
	\If {$b_{j} = 1$} {
		$st \leftarrow$ the likely-buggy statement in the $j$-th modification point\;
		\If {{$u_{j} = 1$}} {
			Delete $st$\;
		}
		\Else {
		$st^* \leftarrow$ the $v_{j}$-th ingredient statement in $I_{j}$\;
			\If {{$u_{j} = 2$}} {
				Replace $st$ with $st^*$\;
			}
		        \ElseIf {$u_{j} = 3$} {
			 	Insert $st^*$ before $st$\;
			}
		}
	}
}
 \caption{The procedure to apply a patch $\mathbf{x}$}
\end{algorithm}

\subsubsection{Population Initialization}
\label{sec-Population Initialization}

For a specific problem, it is usually better to use the initialization strategy based on prior knowledge instead of random initialization, which 
could help genetic search find  desirable solutions more quickly and easily. 

In ARJA, we initialize the first part (i.e., $\mathbf{b}$) of each solution by exploiting the output of fault localization. 
Suppose $susp_{j}$ is the suspiciousness of the likely buggy statement in the $j$-th modification point, then $b_{j}$ is initialized to 1 with 
the probability $susp_{j} \times \mu$ and 0 with $1- susp_{j} \times \mu$, where $\mu \in (0,1)$ is a predefined parameter. 
The remaining two parts (i.e., $\mathbf{u}$ and $\mathbf{v}$) of each solution are just initialized randomly (i.e., $u_{j}$ and $v_{j}$ are initialized to an integer randomly chosen from $[1, |O|]$ and $[1, |I_{j}|]$ respectively).

\subsubsection{Fitness Evaluation}
\label{sec-Fitness Evaluation}

In ARJA, we formulate automated program repair as a multi-objective search problem. 
To evaluate the fitness of a solution $\mathbf{x}$, we propose a multi-objective function
to simultaneously minimize two objectives, namely  \emph{patch size} (denoted by  $f_{1}(\mathbf{x}$)) and  \emph{weighted failure rate} (denoted by  $f_{2}(\mathbf{x}$)).

The patch size is given by Eq. (\ref{eq-f1}), which indeed refers to the number of edit operations contained in the patch. 
\begin{equation}\label{eq-f1}
\begin{aligned}
f_{1}(\mathbf{x}) = \sum_{i=1}^{n}b_{i}
\end{aligned}
\end{equation}

The weighted failure rate measures how well the modified program (obtained by applying the patch $\mathbf{x})$
passes the given tests. We can formulate it as follows:
\begin{equation}\label{eq-f2}
\begin{aligned}
f_{2}(\mathbf{x}) =   \frac{|\{t \in T_{f} \mid \text{$\mathbf{x}$ fails $t$} \}| }{|T_{f}|} + w \times  \frac{ | \{ t \in T_{c} \mid \text{$\mathbf{x}$ fails $t$}\}| } {|T_{c}|}  
\end{aligned}
\end{equation}
where $T_{f}$ is the set of negative tests, $T_{c}$ is the reduced set of positive tests
obtained through test filtering, and $w \in (0,1]$ is a global parameter  which can introduce a bias toward  negative tests.
If $f_{2}(\mathbf{x})=0$,  $\mathbf{x}$ does not fail any test and represents a test-adequate patch.

By simultaneously minimizing $f_{1}$ and $f_{2}$, we prefer  
test-adequate patches of smaller  size. 
Note that if the modified program fails to compile or  runs out of time when executing the tests, we  set both of
the  objectives to $+\infty$. Moreover, $f_{1}=0$ would be meaningless for program repair since no modifications are
made to the orginal program. 
So, once $f_{1}$ is equal to 0 for a solution $\mathbf{x}$, $f_{1}$ and $f_{2}$ are immediately reset to $+\infty$, 
forcing such solutions to disappear with elite selection.

ARJA also provides two optional strategies that are implemented for improving computational efficiency of $f_{1}$
and $f_{2}$ respectively. 
For $f_{1}$, the user can choose to set a threshold $n_{e}$ to avoid disrupting  the original program too much.
Once $n_{e}$  is set beforehand and $f_{1}$ is larger than $n_{e}$ for a solution,  $f_{1}$ is reset to $n_{e}$   and only 
$n_{e}$  edit operations are used on  modification points with the largest suspiciousness  to compute $f_{2}$.
This strategy can also help to accelerate  convergence of the genetic search.

As for $f_{2}$, the strategy is similar to that used by GenProg \cite{le2012systematic}.
That is, if $|T_{c}|$ is still too large,  a random sample of $T_{c}$  (different at each time) can be used for each computation of
$f_{2}$ instead of  $T_{c}$. Only when the resulting $f_{2}$ is 0, all the tests in $T_{c}$ are considered to recalculate $f_{2}$.
The rationale for this strategyis that GP can also work well with a noisy fitness function \cite{fast2010designing}.

\subsubsection{Genetic Operators}
\label{sec-Genetic Operators}

The genetic operators (i.e., crossover and mutation) are executed to
produce  offspring solutions in GP.  
To inherit good traits from  parents,  crossover and mutation are applied to each part of the 
solution representation separately. 

\begin{figure}[htbp]
\centering
        \includegraphics[scale=0.52]{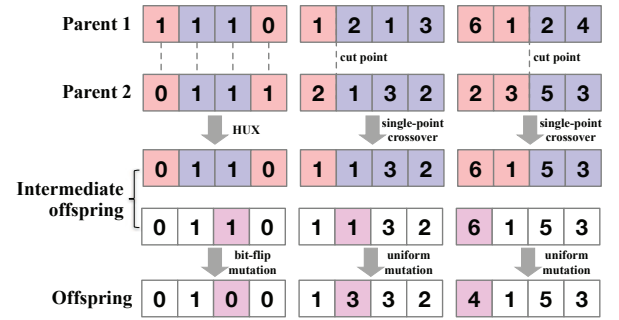}
\caption{Illustration of crossover and mutation in ARJA.} \label{fig-go}
\end{figure}

For the first part (i.e., $\mathbf{b}$), we use  half uniform crossover (HUX)
and  bit-flip mutation. 
As for both of the  remaining parts (i.e.,  $\mathbf{u}$ and $\mathbf{v}$), we adopt  single-point crossover and 
 uniform mutation, because of their integer encoding. 
Fig. \ref{fig-go} illustrates how  crossover and mutation are executed on
two parent solutions. For brevity, only one offspring is shown in this figure. 

\begin{table*}[htbp]
\renewcommand{\arraystretch}{1.0}
\centering
\caption{The rules integrated in ARJA for customizing the operation types for each modification point}
\label{tab-otrule} \footnotesize \tabcolsep =15pt
\begin{tabular}{lll}
\toprule
No.   & Rule    &     rationale\\
\midrule
1   &   Do not delete a variable declaration statement (VDS).  & Deleting a VDS is usually very disruptive to a program, \\
      &                                                                                              & and keeping a redundant VDS usually does not influence  \\
      	&										        & the correctness of a program.  \\
 \cmidrule{1-3}
2   &   Do not delete a \lstinline[language=Java]!return!/\lstinline[language=Java]!throw! statement which  & Avoid returning no value from a method that is not\\
    & is the last statement of a method not declared \lstinline[language=Java]!void!. &declared \lstinline[language=Java]!void!. \\
\bottomrule
\end{tabular}
\end{table*}

\begin{table*}[htbp]
\renewcommand{\arraystretch}{1.0}
\centering
\caption{The rules integrated in ARJA for further filtering the ingredients for each modification point}
\label{tab-ingrule} \footnotesize \tabcolsep =15pt
\begin{tabular}{lll}
\toprule
No.   & Rule    &     rationale \\
\midrule
1   & The \lstinline[language=Java]!continue! statement can be used as the & The keyword \lstinline[language=Java]!continue! cannot be used out of a loop		\\					       
	& ingredient only for a likely-buggy statement   &   (i.e., \lstinline[language=Java]!for!, \lstinline[language=Java]!while! or \lstinline[language=Java]!do-while! loop).  \\
     & in the loop.  \\
 \cmidrule{1-3}
2   &The \lstinline[language=Java]!break! statement can be used as the & The keyword \lstinline[language=Java]!break! cannot be used out of a loop \\
     &ingredient only for a likely-buggy statement in   &  (i.e., \lstinline[language=Java]!for!, \lstinline[language=Java]!while! or \lstinline[language=Java]!do-while! loop) or a \lstinline[language=Java]!switch! \\
     &the loop or in the \lstinline[language=Java]!switch! block. &     block. \\
 \cmidrule{1-3}

 3 & A \lstinline[language=Java]!case! statement can be used as the ingredient   & The keyworkd  \lstinline[language=Java]!case! cannot be used out of a \lstinline[language=Java]!switch! \\
   &only for a likely-buggy statement in a \lstinline[language=Java]!switch!   &  block, and the value for a \lstinline[language=Java]!case! must be the same \\
   &block having the same enumerated type. &  enumerated type as the variable in the \lstinline[language=Java]!switch!. \\
   
   \cmidrule{1-3}
   
4 & A \lstinline[language=Java]!return!/ \lstinline[language=Java]!throw! statement can be used as the  & Avoid returning/throwing a value with non-compatible \\
   & ingredient only for a likely-buggy statement in a    & type from a method.  \\
   &method declaring the compatible return/throw type. \\
 \cmidrule{1-3}   
 
5 & A \lstinline[language=Java]!return!/ \lstinline[language=Java]!throw! statement can be used as the  & Avoid the unreachable statements. \\
   & ingredient only for a likely-buggy statement that is    &  \\
   & the last statement of a block.\\
 \cmidrule{1-3}   
   
6 & A VDS can be used as the ingredient only for another   & Avoid using an edit operation with no effect on the \\
   & VDS having the compatible declared type and the  		& program or disrupting the program too much.   \\
   &  same variable names. \\   
\bottomrule
\end{tabular}
\end{table*}

\subsubsection{Using NSGA-II}
\label{sec-Using NSGA-II}

Generally, based on the proposed solution representation, any MOEA can 
serve the purpose of evolving the patches for a buggy program. In ARJA, we employ NSGA-II \cite{deb2002fast}
as the  search algorithm, which is one of the most popular MOEA. 

The NSGA-II based search procedure for finding  test-adequate patches can be summarized as follows.
First, an initial population with $N$ (the population size) solutions is produced by
using the initialization strategy presented in Section \ref{sec-Population Initialization}. Then the algorithm goes into
a loop until the maximum number of fitness evaluations is reached. In each generation $g$, binary tournament selection \cite{deb2002fast} and the genetic operators described in Section \ref{sec-Genetic Operators}
are applied to the current population $P_{g}$ to generate
an offspring population $Q_{g}$. Then the $N$  best solutions 
are selected from the union population
$U_{g} = P_{g} \cup Q_{g}$ by using fast non-dominated sorting and crowding distance
comparison (based on the two objectives formulated in Section \ref{sec-Fitness Evaluation}). 
The resulting $N$ best solutions constitute the next population $P_{g+1}$.

Finally, the obtained non-dominated solutions with $f_{2}=0$
are output as  test-adequate patches found by ARJA. If 
no such solutions exist, ARJA fails to fix the bug.

\subsection{Rule-Based Search Space Reduction}
\label{sec-Rule-Based Search Space Reduction}

\begin{table*}[htbp]
\renewcommand{\arraystretch}{1.0}
\centering
\caption{The rules integrated in ARJA for disabling certain specific operations}
\label{tab-oirule} \footnotesize \tabcolsep =15pt
\begin{tabular}{lll}
\toprule
No.   & Rule    &     rationale \\
\midrule
1  & Do not replace a statement with the one having& Avoid using an edit operation with no effect   \\

  & 	 the same AST.	&on the program. \\
 \cmidrule{1-3}
 
2   & Do not replace a VDS with the other kinds of statements. & Avoid disrupting the program too much. \\
  \cmidrule{1-3}
  
3 & Do not insert a VDS before a VDS.  & The same with No. 1.\\
   \cmidrule{1-3}
 
4 & Do not insert a \lstinline[language=Java]!return!/\lstinline[language=Java]!throw! statement before any  & Avoid the unreachable statements. \\
   & statement. \\
   \cmidrule{1-3}

5 & Do not replace a \lstinline[language=Java]!return! statement (with return value) that is  & Avoid returning no value from a method  \\
 & the last statement of a method with the other kinds of statements. &  that is not declared void.\\
    \cmidrule{1-3} 

6 &  Do not insert an assignment statement before an   & The same with No. 1.\\
   & assignment statement with the same left-hand side. & \\
\bottomrule
\end{tabular}
\end{table*}

ARJA provides three types of rules that can be integrated into its three different procedures (i.e., operation initialization, ingredient screening and solution decoding)
respectively. By taking advantage of these rules, we can not only increase the chance of the modified program 
to compile successfully , but also avoid some meaningless edit operations,
thereby reducing the search space. 

Note that when the rules are integrated into ARJA, the related procedures described in the previous subsections will be 
modified as discussed next.

\subsubsection{Customizing the Operation Types}
\label{sec-Customizing the Operation Types}

The first type of rules are used to customize the operation types for each modification point.  Such rules are invoked in the operation initialization procedure since they only involve 
 likely-buggy statements and the operation types.

\begin{figure}[htbp]
\centering
        \includegraphics[scale=0.45]{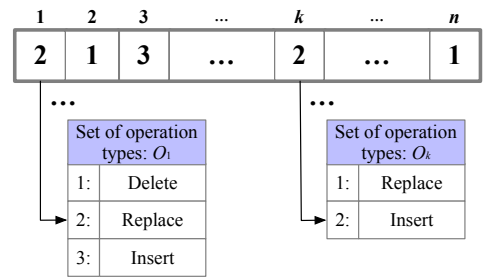}
\caption{Illustration of $\mathbf{u}$ for the purpose of customizing the operation types.} \label{fig-uv_rule}
\end{figure}

For a modification point, certain operation types in $O$ may not be available according to 
the predefined rules. Currently in ARJA, we provide two rules of this type which are shown in 
Table \ref{tab-otrule}. Suppose $O_{j}$ is the set of available operation types for the $j$-th modification point, 
where $O_{j} \subseteq O, j=1,2,\ldots, n$. 
Fig. \ref{fig-uv_rule} illustrates the $\mathbf{u}$ vector for the purpose of customizing the operation types.
Unlike in Fig \ref{fig-uv}, each modification point is associated with its own set of operation types (i.e., $O_{j}$),
and $u_{j}$ means that the patch chooses the $u_{j}$-th operation type in $O_{j}$ (instead of $O$).

\subsubsection{Further Filtering the Ingredients}
\label{sec-Further Filtering the Ingredients}

The second type of rules  concerns  likely-buggy statements and 
the ingredient statements, which are employed to further filter the 
ingredient statements in an ingredient screening procedure.
Such rules can help to remove  undesirable ingredients which 
pass the scope check of variables and methods. 
For example, a \lstinline[language=Java]!continue! statement 
does not contain any variable or method invocation, but it can only be used
in a loop. Table \ref{tab-ingrule} lists the rules of this type integrated into ARJA, and
also explains their rationale. 

By applying these rules to $I_{j}$ (obtained by the procedure in Section \ref{sec-Ingredient Screening}), we can generate 
a reduced set $I_{j}'$ where $I_{j}'\subseteq I_{j}$. 
$I_{j}'$ will become the set of ingredients for the $j$-th modification point instead of $I_{j}$.
To illustrate $\mathbf{v}$ in this scenario, we can just replace $I_{j}$ with $I_{j}'$ in Fig. \ref{fig-vv}, with $v_{j}$ then indicating
that the patch chooses the $v_{j}$-th ingredient statement in $I_{j}'$.

\subsubsection{Disabling Certain Specific Operations}
\label{sec-Disabling Certain Specific Operations}

The third type of rules involve at least the
operation types and the ingredient statements. Such rules 
are used to ignore certain specific edit operations when 
decoding a solution $\mathbf{x}$ for fitness evaluation.  
Table \ref{tab-oirule} shows the rules of
this type integrated in ARJA together with their rationale.

With these rules, if $b_{j}=1$, the corresponding operation on the $j$-th modification point will not be conducted immediately as in Algorithm \ref{alg-patch}.
Instead, we first check whether this operation conforms to one rule listed in Table \ref{tab-oirule}. 
Once any of the rules is met, the operation will be disabled (equivalent to resetting $b_{j}$ to 0).

\section{Experimental Design}
\label{sec-Experimental Design}

This section explains the design of our experimental study, including
the research questions to be answered, the repair systems involved, the
datasets of bugs used, and the evaluation protocol for comparing different repair approaches. 

\subsection{Research Questions}
\label{sec-Research Questions}

To conduct the general evaluation of ARJA, we seek to answer the following research questions in
this study. 

\begin{itemize}[leftmargin=0mm]
 \item[] \textbf{RQ1:} How useful is the test filtering procedure in speeding up the fitness evaluation?
\end{itemize}

In this research question, we show what percentage of CPU time for fitness evaluation can be saved by test filtering.

\begin{itemize}[leftmargin=0mm]
 \item[] \textbf{RQ2:} Does random search really outperform genetic search in automated program repair?
\end{itemize}

Previous work by Qi et al. \cite{qi2014strength} claimed that
random search outperforms GP in terms of both repair effectiveness and efficiency. Their work targeted
C programs and was based on GenProg framework. We are interested in revisiting 
this claim based on ARJA that targets Java programs. 

\begin{itemize}[leftmargin=0mm]
 \item[] \textbf{RQ3:} What are the benefits of formulating program repair as a multi-objective search problem?
\end{itemize}

We would expect that the multi-objective formulation described in Section \ref{sec-Fitness Evaluation}
can help ARJA generate simpler patches compared to a single-objective formulation. Besides this, 
we investigate whether the multi-objective formulation can provide other benefits.

\begin{itemize}[leftmargin=0mm]
 \item[] \textbf{RQ4:}  Is ARJA better than state-of-the-art repair methods in terms of fixing multi-location bugs?
\end{itemize}

As stated in Section \ref{sec-Introduction}, one prominent feature of GP based repair approaches is that they have the  potential to
fix multi-location bugs. ARJA is a new GP based repair system, thus it 
is necessary to assess its superiority in this respect.

In RQs 2--4, our main concern is the \emph{search ability}
of ARJA (including its variants) and other related repair
methods based on the redundancy assumption.
So, here we only use ARJA \emph{without} type matching for the simplicity and fair comparison.


\begin{itemize}[leftmargin=0mm]
 \item[] \textbf{RQ5:} How useful is the type matching strategy when the fix ingredients do not exist in the current buggy program?
\end{itemize}

Type matching can reasonably create ingredient statements that do not appear in the buggy program. 
We investigate whether these newly generated ingredients can be exploited effectively by 
ARJA to fix some bugs.

\begin{itemize}[leftmargin=0mm]
 \item[] \textbf{RQ6:} How well does ARJA perform in fixing real-world bugs compared to the existing repair approaches?
 \end{itemize}
 
It is of major interest to address real-world bugs in  program repair. 
We need to know whether ARJA can work on real-world bugs in large-scale Java software systems, beyond fixing seeded bugs.

\begin{itemize}[leftmargin=0mm]
 \item[] \textbf{RQ7:} To what extent can ARJA synthesize semantically correct patches?
\end{itemize}

Since the empirical work by Qi et al. \cite{qi2015analysis}, it has been a hot question whether
the patches generated by test-suite based repair approaches
are correct beyond passing the test suite. We manually check the correctness of the patches 
synthesized by ARJA in our experiments. 
 
 \begin{itemize}[leftmargin=0mm]
 \item[] \textbf{RQ8:} Why can't ARJA generate test-suite adequate patches for some bugs?
 \end{itemize}

Sometimes,  ARJA  fails to find any test-suite adequate patch. We examine several reasons for failure. 
 
\begin{itemize}[leftmargin=0mm]
 \item[] \textbf{RQ9:} How long is the execution time for ARJA on one bug?
\end{itemize}

The computational cost of one repair is also an important concern for users. We
want to see whether the repair cost of ARJA is acceptable for practical use. 
 
\subsection{Repair Systems Under Investigation}
\label{sec-Repair Systems Under Investigation}

There are mainly five repair systems involved in our experiments, which are ARJA,
GenProg \cite{le2012systematic}, RSRepair \cite{qi2014strength}, Kali \cite{qi2015analysis} and Nopol \cite{xuan2017nopol}. 

Our repair system, ARJA, is implemented with Java 1.7 on top of
jMetal 4.5.\footnote{jMetal, http://jmetal.sourceforge.net} jMetal \cite{durillo2011jmetal} is a 
Java based framework that includes a number of state-of-the-art EAs, particularly MOEAs. 
It is used to provide computational search algorithms (e.g., NSGA-II)
in our work. ARJA parses and manipulates Java source code 
using the Eclipse JDT Core\footnote{Eclipse JDK Core, https://www.eclipse.org/jdt/core/index.php} package.  
The fault localization in ARJA is implemented with Gzoltar 0.1.1.\footnote{Gzoltar, http://www.gzoltar.com. The version 0.1.1 cannot localize the faults in a constructor. So when 
repairing some bugs in Defects4J that are located in a constructor, we switch to the version 1.6.2 although the new version appears much more computationally intensive}
Gzoltar \cite{campos2012gzoltar} is a toolset which determines
suspiciousness of faulty statements using 
spectrum-based fault localization algorithms.
Both coverage analysis and test filtering in ARJA are implemented with JaCoCo 0.7.9,\footnote{JaCoCo, http://www.eclemma.org/jacoco}
which is a  Java code coverage library. 
For the sake of reproducible research, the source code of 
ARJA is available at GitHub.\footnote{ARJA, https://github.com/yyxhdy/arja}
In addition, several ARJA variants
have also been implemented to answer different research questions. 

RSRepair is a repair method that always uses
the population initialization procedure of GenProg to 
produce candidate patches. Kali generates a
patch by just removing or skipping statements. 
Strictly speaking, Kali cannot be regarded as
a ``program repair'' technique, but it is
a very suitable technique for identifying 
weak test suites or under-specified bugs \cite{martinez2016automatic}. 
GenProg, RSRepair, and Kali were originally developed for C programs. 
According to the details given in the corresponding papers \cite{le2012systematic,qi2014strength,xuan2017nopol},
we carefully reimplement the three systems for Java under the same infrastructure of
ARJA. Our source code for the three systems is publicly released along with ARJA.\footnotemark[\value{footnote}]
Note that a program repair library named Astor \cite{martinez2016astor} also provides the implementation of GenProg and Kali for Java, which are
called jGenProg and jKali respectively in the literature \cite{martinez2016astor,martinez2016automatic}. 
But to conduct controlled experiments, we only use our own implementation, unless otherwise specified.

Nopol is a state-of-the-art semantic-based repair method for fixing conditional
statement bugs in Java programs. The code of Nopol\footnote{Nopol, https://github.com/SpoonLabs/nopol} has been 
released by the original authors.

\subsection{Datasets of Bugs}
\label{sec-Datasets of Bugs}

In our experiments, we use both seeded bugs and real-world bugs to
evaluate the performance of repair systems. 

To answer RQ1 and RQs 6--9, we adopt a dataset
consisting of four open-source Java projects (i.e., JFreeChart,\footnote{JFreeChart, http://www.jfree.org/jfreechart} Joda-Time,\footnote{Joda-Time, http://www.joda.org/joda-time} Commons Lang, and Commons Math) from Defects4J \cite{just2014defects4j}. 
Defects4J\footnote{Defects4J, https://github.com/rjust/defects4j} has been a popular database for evaluating Java program repair 
systems \cite{martinez2016automatic, le2016history, xiong2017precise, yu2017test, xin2017identifying}, 
because it contains well-organized real-world Java bugs. 
Table \ref{tab-d4j}
shows the basic information of the 224 real-world bugs considered in Defects4J, 
where the number of lines of code and the number of JUnit tests are 
extracted from the latest buggy version of each project. 
Note that Defects4J indeed contains another two projects, namely 
Closure Compiler\footnote{Closure Compiler, https://github.com/google/closure-compiler} and Mockito\footnote{Mockito, http://site.mockito.org}.
Following the practice in refs. \cite{martinez2016automatic, xiong2017precise, yu2017test}, we do not
consider the two projects in the experiments. Closure Compiler is dropped since it
uses the customized testing format rather than the standard JUnit tests; Mockito is ignored
because it is a very recent project added into the Defects4J framework and its related artifacts
are still in an unstable phase. 

\begin{table}[htbp]
\renewcommand{\arraystretch}{1.0}
\centering
\caption{The descriptive statistics of 224 bugs considered in Defects4J}
\label{tab-d4j} \footnotesize \tabcolsep =5pt
\begin{tabular}{cccccc}
\toprule
\multirow{2}{*}{Project}  &  \multirow{2}{*}{ID}          &  \multirow{2}{*}{\#Bugs}   & \multirow{2}{*}{\#JUnit Tests}  &Source  & Test  \\
             &                &                 &                          & KLoC    & KLoC      \\
\midrule
JFreeChart           & C  & 26     & 2,205  &96  & 50\\
Joda-Time            & T  & 27     &4,043    &28  & 53 \\
Commons Lang   & L    & 65    &  2,295  &22 & 6\\
Commons Math   & M   & 106 & 5,246    &85  & 19 \\
\cmidrule{1-6}
Total     &          & 224       &  13,789                    & 231 & 128  \\
\bottomrule
\end{tabular}
\end{table}

To address RQs 2--4, we use a dataset of seeded bugs rather than 
Defects4J. We think that Defects4J is not well suited to the purpose of
distinguishing clearly the search ability of the  repair systems considered. The reasons are listed as follows:
\begin{enumerate}
\item Many bugs (e.g., C2, L4 and M12) in Defects4J cannot be localized by 
state-of-the-art fault localization tools (e.g., Gzoltar).
In such a case, fault localization rather than the search is responsible for the failure.
\item Although  existing empirical studies \cite{martinez2014fix,barr2014plastic}
validated the redundancy assumption by examining a large number of commits (16,071 in ref. \cite{martinez2014fix} and 15,723 in ref. \cite{barr2014plastic}), 
Defects4J contains a relatively small number of bugs and it may not conform well
to this general assumption. Indirect evidence is that 
jGenProg can find patches for only 27 out of 224 bugs in Defects4J, as reported by Martinez et al. \cite{martinez2016automatic}. 
In such a case, it is the inadequate search space that matters, rather than the search ability. 
\item As indicated in ref. \cite{martinez2016automatic}, among 27 bugs fixed by GenProg, 
20 bugs can also be fixed by jKali. This means that for the 
overwhelming majority of these bugs,
the search method 
can find a trivial patch (e.g., deleting a statement)
that fulfills the test-suite by just focusing on a very limited search space.
So,  evaluation on such bugs cannot truly reflect the difference between  redundancy-based repair methods in exploring a
huge search space of potential fix ingredients. 
\end{enumerate}

The dataset of seeded bugs for RQs 2--4 are generated by the following procedures. 
First, we select the correct version of M85 as a target program, since it 
has a moderate number (i.e., 1983) of JUnit tests. Then, we randomly
select $k$ redundant\footnote{Here ``redundant'' means that the same statement can be found otherwhere in the current package.} statements 
from two Java files (i.e., NormalDistributionImpl.java and PascalDistributionImpl.java). Last, 
we produce a buggy program by performing mutation to each of the $k$ statements. 
Note that not every buggy program obtained in this way is a suitable test bed, we
choose some of them according to the following principles:
\begin{enumerate}
\item The fault localization technique can identify all the faulty locations of
the seeded bug. This rules out the influence of fault localization. 
\item Any nonempty subset of the $k$ mutations should make at least one test fail. Generally, this
ensures that the seeded bug is really a multi-location bug when $k >1$. 
\item Kali cannot generate any test-suite adequate patch for the seeded bug. This challenges
the search ability in finding nontrivial or complex repairs. 
\end{enumerate}

We vary $k$ from 1 to 3 and finally collect a dataset consisting of 13 bugs of this kind, denoted by F1--F13. 
Among the 13 bugs, $k$ is set to 1 for F1 and F2, 2 for F3--F9, and 3 for F10--F13. So, all bugs except F1 and F2
are multi-location bugs.  
Because the mutated statements are redundant, the redundancy-based repair
systems (e.g., GenProg and RSRepair) can fix any bug in this dataset, assuming
that their search ability is strong enough. 

For RQ5, we use a similar method
to generate a dataset of seeded bugs. 
The difference is that among the $k$ statements
to be mutated, at least one is non-redundant. 
So, it is expected that the redundancy assumption does not completely hold for
such bugs, which can be used to verify the effectiveness of
the type matching strategy. We set $k$ to 2 and collect 5 bugs in this category, denoted by H1--H5. 

To facilitate experimental reproducibility,
we make the two datasets of seeded bugs available on GitHub as well.\footnote{Seeded bugs, http://github.com/yyxhdy/SeededBugs}



\subsection{Evaluation Protocol}
\label{sec-Evaluation Protocol}

In our experiments, we always use ``Package'' as the ingredient mode
in ARJA, GenProg, and RSRepair, although there exist  two other
alternatives as introduced in Section \ref{sec-Ingredient Screening}. 
To reduce the search space, we integrate all three types of rules (see Section \ref{sec-Rule-Based Search Space Reduction}) into ARJA. 
When investigating RQs 2--4, 
the direct approach (see Section \ref{sec-Direct Method}) is used in ARJA for ingredient screening.\footnote{Hereafter, if ``ARJA'' represents a specific algorithm, it refers to
the version that uses the direct approach for ingredient screening. The ARJA variants using type matching will be differentiated by subscripts}
To save  computational time, we employ the 
test filtering procedure (see Section \ref{sec-Test Filtering}) in all  repair approaches implemented by ourselves (i.e., ARJA, GenProg, RSRepair, and Kali).

\begin{table}[htbp]
\renewcommand{\arraystretch}{1.0}
\centering
\caption{The parameter setting for ARJA in the experiments}
\label{tab-parameters} \footnotesize \tabcolsep =5pt
\begin{tabular}{clc}
\toprule
Parameter  & Description        &   Value\\
\midrule
$N$   & Population size      &  40 \\
$G$  & Maximum number of generations & 50 \\
$\gamma_{\min}$ &Threshold for the suspiciousness     & 0.1\\
$n_{\max}$ & Maximum number of modification points  & 40 \\
$\mu$ & Refer to Section \ref{sec-Population Initialization} & 0.06\\
$w$ & Refer to Section \ref{sec-Fitness Evaluation} & 0.5\\
$p_{c}$ & Crossover probability & 1.0\\
$p_{m}$ &Mutation probability &  $1 / n$\\
\bottomrule
\end{tabular}
\end{table}

Table \ref{tab-parameters} presents the basic parameter setting for ARJA in the experimental study, where $n$
is the number of modification points determined by $\gamma_{\min}$ and $n_{\max}$ together (see Section \ref{sec-Fault Localization and Coverage Analysis}).
To ensure a fair comparison,  parameters $N$, $G$, $\gamma_{\min}$ and $n_{\max}$ in GenProg and RSRepair are set to the same values as shown in Table \ref{tab-parameters}. 
The global mutation rate in GenProg and RSRepair is set to 0.06 since it is similar to  parameter $\mu$ in ARJA. 
Corresponding to $w=0.5$,  negative tests are weighted twice as heavily as positive tests in  fitness evaluation of GenProg and RSRepair.  
Each trial of ARJA, GenProg and RSRepair is terminated after the maximum number of generations is reached. 
Moreover, in Kali, we also use $\gamma_{\min}$ and $n_{\max}$ to restrict the number of modification points considered.
Their values are set to 0.1 and 40 respectively.

ARJA, GenProg, and RSRepair are all stochastic search methods. 
To compare their search ability properly, each of these algorithms 
performs 30 independent trials for each seeded bug considered in RQs 2--5. 
There are several metrics involved for evaluating the search performance, which
are explained as follows:
\begin{enumerate}
\item ``Success'': the number of trials that produce at least one test-suite adequate patch among 30 independent trials. This is regarded as the primary metric. 
\item ``\#Evaluations'' and ``CPU (s)'': the average number of evaluations and the average CPU time needed to find the first test-suite adequate patch in a successful trial. 
\item ``Patch Size'': the average size of the smallest test-suite adequate patch obtained in a successful trial. Here  ``size'' means
the number of edits contained in a patch. 
\item ``\#Patches'': the average number of different test-suite adequate patches found in a successful trial.
Note that we may obtain test-suite adequate patches with various sizes in a trial,
this metric only counts the number of those with the smallest size. Moreover, the difference 
between patches here is judged in terms of syntactics rather than semantics. 
\end{enumerate}

Following the practice in ref. \cite{martinez2016automatic}, we 
perform only one trial of ARJA, GenProg, and RSRepair for each real-world bug considered in RQ6,
in order to keep  experimental time acceptable. It is indeed very CPU intensive to conduct a large-scale 
experiment on the 224 bugs from Defects4J. For example, Martinez et al. \cite{martinez2016automatic}
ran each of three algorithms (i.e., jGenProg, jKali and Nopol) on each of the 224 bugs  only once, which
took up to 17.6 days of computational time on Grid'5000. However, we note that
multiple trials are needed to rigorously assess the performance of ARJA, GenProg, and RSRepair
due to their stochastic nature. We discuss this important threat to validity in Section \ref{sec-Threats to Validity}.

Our experiments were all performed in the MSU High Performance Computing Center.\footnote{MSU HPCC, https://wiki.hpcc.msu.edu/display/hpccdocs}
We use 2.4 GHz Intel Xeon E5 processor machines with 20 GB memory.

\section{Experimental Results and Analysis}
\label{sec-Experimental Results and Analysis}

This section presents the results of our experimental study in order to 
address RQs 1--9 set out in Section \ref{sec-Research Questions}.

\subsection{Value of Test Filtering (RQ1)}
\label{sec-Value of Test Filtering (RQ1)}

To answer RQ1, we select the latest buggy versions of
the four projects considered in Defects4J (i.e., C1, T1, L1 and M1) as the subject programs. 

\begin{figure}[htbp]
\centering
        \includegraphics[scale=0.46]{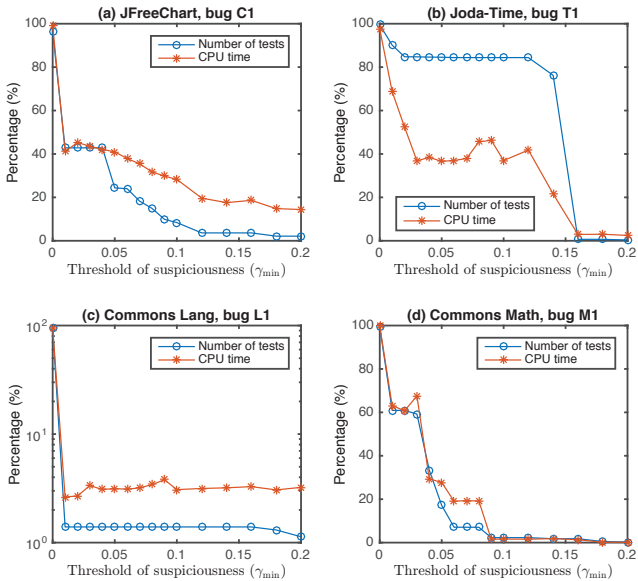}
\caption{Illustration of the value of the test filtering procedure. The base 10 logarithmic scale is used for the $y$ axis in (c).} \label{fig-vtf}
\end{figure}

For each buggy program, 
we vary $\gamma_{\min}$ (i.e., the threshold of suspiciousness)
from 0 to 0.2. For each $\gamma_{\min}$ value chosen,
we use the test filtering procedure to obtain a subset of original JUnit tests
and then record two metrics associated with the reduced test suite: the number of
tests and the execution time. 
Fig. \ref{fig-vtf} shows the influence of $\gamma_{\min}$ on the two metrics for each subject program. 
For comparison purposes, the two metrics have been normalized by dividing 
by the original number of JUnit tests and the CPU time consumed by the original test suite respectively. 
Note that the fluctuation of CPU time in Fig. \ref{fig-vtf}  is due to the dynamic behavior of the computer system.

Judged from Fig. \ref{fig-vtf}, test filtering can bring substantial benefits 
in terms of reduction of computational costs. With the increase in $\gamma_{\min}$
the number of tests that need to be considered
and the corresponding CPU time consumed decrease significantly and quickly.
Even if we set $\gamma_{\min}$ to a very small value (e.g., 0.01), 
test filtering can result in a considerable reduction of CPU time. Suppose $\gamma_{\min}=0.01$, the percentage 
reduction is about 59\% for C1, 31\% for  T1, 97\% for L1 and 37\% for M1. 
Generally, if we set $\gamma_{\min}$ to a larger value, we can consider a smaller test suite ifor fitness evaluation.
However, it is not desirable to use  too large a $\gamma_{\min}$ (e.g., 0.5), which would make 
 repair approaches miss the actual faulty locations. 
In practice, we usually choose a moderate $\gamma_{\min}$ value (e.g., 0.1) to strike a compromise.
Normally, test filtering can significantly speed up the fitness evaluation in such a case. For example, 
if we set $\gamma_{\min}$ to 0.1 for M1, the number of tests considered is reduced from 5,246 to 118, and 
the CPU time for one fitness evaluation is reduced from 210 seconds to 3.4 seconds. Suppose the termination
criterion of ARJA is 2,000 evaluations, then we can save up to 115 hours for just a single repair trial. 


For safety, we also conduct a post-run validation of the obtained patches on the original test suite.
We find that if a patch can fulfill the reduced test suite, it can also fulfill the original one, with no exceptions.

\subsection{Genetic Search vs. Random Search (RQ2)}
\label{sec-Genetic Search vs. Random Search (RQ2)}

To compare the performance of genetic search and random search 
under the ARJA framework, we implemented an ARJA variant denoted as 
$\text{ARJA}_{r}$.
The only difference between ARJA and $\text{ARJA}_{r}$ lies in that $\text{ARJA}_{r}$ always uses the initialization procedure of ARJA to 
generate  candidate solutions and does not use the genetic operators described in 
Section \ref{sec-Genetic Operators}. So, $\text{ARJA}_{r}$ purely depends on the random search and there is no cumulative selection. 
The relationship between ARJA and $\text{ARJA}_{r}$ is similar to that between GenProg and RSRepair. 
For a fair comparison, $\text{ARJA}_{r}$ also uses the parameter setting shown in Table \ref{tab-parameters} (excluding $p_{c}$ and $p_{m}$).

Table \ref{tab-gr} compares ARJA and $\text{ARJA}_{r}$ on F1--F13 in terms of the metrics ``Success'' and ``\#Evaluations''.
In this table, the meaning of $k$ can be referred to in Section \ref{sec-Datasets of Bugs} and $|T_{f}|$ is the number of negative tests that trigger the bug.
For brevity, the two columns will be omitted later in Tables \ref{tab-ms} and \ref{tab-sfm}.

\begin{table}[htbp]
\renewcommand{\arraystretch}{1.0}
\centering
\begin{threeparttable}
\caption{Comparison between genetic search and random search within the ARJA framework. (Average over 30 runs)}
\label{tab-gr} \footnotesize \tabcolsep =5pt
\begin{tabular}{cccccccc}
\toprule
\multirow{2}{2.5em}{Bug\\Index}  &  \multirow{2}{*}{$k$} &  \multirow{2}{*}{$|T_{f}|$} & \multicolumn{2}{c}{Success}   &&   \multicolumn{2}{c}{\#Evaluations} \\
                                                                                          \cmidrule{4-5} \cmidrule{7-8}
                                                     &                                  &                                           &   ARJA  & $\text{ARJA}_{r}$      &&  ARJA  & $\text{ARJA}_{r}$          \\
\midrule
F1 & 1 & 3 & 30 & 10 && 297.67 & 507.60\\
F2 & 1 & 4 & 17 & 19 && 492.71 & 392.32\\
F3 & 2 & 4 & 26 & 3 && 494.24 & 668.00\\
F4 & 2 & 6 & 13 & 0 && 746.54 & --\\
F5 & 2 & 8 & 30 & 3 && 384.63 & 111.00\\
F6 & 2 & 4 & 30 & 1 && 624.80 & 1229.00\\
F7 & 2 & 3 & 25 & 0 && 698.52 & --\\
F8 & 2 & 6 & 29 & 4 && 376.21 & 505.50\\
F9 & 2 & 2 & 6 & 0 && 1028.00 & --\\
F10 & 3 & 6 & 18 & 0 && 936.11 & --\\
F11 & 3 & 6 & 20 & 0 && 777.70 & --\\
F12 & 3 & 8 & 28 & 0 && 742.04 & --\\
F13 & 3 & 4 & 20 & 0 && 762.30 & --\\
\bottomrule
\end{tabular}
\begin{tablenotes}
        \item ``--'' means the data is not available. 
\end{tablenotes}
\end{threeparttable}
\end{table}

As can be seen from Table \ref{tab-gr}, on all the  bugs considered except F2 and F5, ARJA achieves a much higher success rate and also 
requires less number of evaluations to find a repair patch compared to $\text{ARJA}_{r}$. 
Moreover, ARJA is much more effective than $\text{ARJA}_{r}$ in synthesizing 
multi-line patches. For example, on each of F10--F13 which need
at least three edit operations, $\text{ARJA}_{r}$ cannot
find any test-suite adequate patch in any of the 30 trials, whereas ARJA still maintains  good performance
and succeeds in the majority of trials. For the bug F5, $\text{ARJA}_{r}$ appears
more efficient since it can find a repair more quickly, but its repair success rate is very low (3 out of 30) and
it is therefore not reliable. 
In contrast to $\text{ARJA}_{r}$, ARJA can always succeed in fixing the bug F5. As for F2, 
$\text{ARJA}_{r}$ performs slightly better than ARJA. 
The possible reason is that the fix of F2 only requires one insertion operation and $\text{ARJA}_{r}$  focuses
more on a search space containing such simple repairs.

In summary, ARJA significantly outperforms $\text{ARJA}_{r}$ in terms of both repair effectiveness
and efficiency, particularly on multi-location bugs, which
indicates that genetic search is indeed more powerful than random search in
automated program repair. 

Note that our conclusion here  contradicts that drawn by Qi et al. \cite{qi2014strength}.
This can be attributed to the fact that the two studies are
based on different algorithmic frameworks and
different subject programs. In ref. \cite{qi2014strength}, 
GenProg and RSRepair were compared on 24 C bugs 
from the GenProg benchmark. As pointed out in ref. \cite{qi2015analysis}, almost
all the patches reported by GenProg and RSRepair for
these 24 bugs are equivalent to a single functionality deletion modification. 
When searching such trivial repairs, 
the crossover operator in GenProg will become ineffective. 
The main reason is that GenProg crossover works on the granularity of edits (as mentioned in Section \ref{sec-High-Granularity Patch Representation}) and
produces a new patch just by combining the edits from two parent solutions without creating any new material. But it is clear that the recombination of  existing edits will not be
helpful to find a patch that contains only a single edit. In addition, because RSRepair always uses
the initialization procedure of GenProg, it has a very high 
chance to generate patches with only one edit when using
a small global mutation rate (e.g., 0.06). Whereas in GenProg, the new
edits will be appended to the existing patch, which means that GenProg intends to
explore larger patches  during the search. Nevertheless, this search characteristic of GenProg may
make it less efficient than RSRepair in finding trivial patches that are test-suite adequate for the bugs considered in ref. \cite{qi2014strength}. We 
speculate that GenProg will outperform RSRepair in terms of generating nontrivial or complex repairs.

\subsection{Multi-Objective vs. Single-Objective (RQ3)}
\label{sec-Multi-Objective vs. Single-Objective (RQ3)}
To show the benefits of the multi-objective formulation, 
we develop an ARJA variant denoted as $\text{ARJA}_{s}$, 
which only minimizes the weighted failure rate (see Eq. (\ref{eq-f2})). 
To serve the purpose of single-objective optimization, 
$\text{ARJA}_{s}$ uses a canonical single-objective GA instead of NSGA-II
to evolve  patches. To ensure a fair comparison, the single-objective GA also 
employs binary tournament selection and the genetic operators introduced in Section \ref{sec-Genetic Operators}, 
and the parameter setting of $\text{ARJA}_{s}$ is the same with that of ARJA.

In Table \ref{tab-ms}, we present the comparative results between ARJA and $\text{ARJA}_{s}$ on bugs F1--F13, where
the metrics ``Success'', ``Patch Size'' and ``\#Patches'' are considered. 
As expected, ARJA can really generate test-suite adequate patches that contain
smaller number of edits. The only exception is F5, where the patch sizes obtained by
ARJA and $\text{ARJA}_{s}$ have no obvious difference. Moreover, it can be
seen that the average patch size obtained by ARJA is usually very
close to the corresponding $k$ value (see Table \ref{tab-gr}) of the bug,
demonstrating the effective minimization of $f_{1}$ (see Eq. (\ref{eq-f1})) by NSGA-II.
According to ``\#Patches'', for every bug, ARJA can find notably more different test-suite adequate patches than $\text{ARJA}_{s}$
in a successful trial, which is expected to provide more adequate choice for the programmer. 

\begin{table}[htbp]
\renewcommand{\arraystretch}{1.0}
\centering
\begin{threeparttable}
\caption{Comparison between multi-objective and single-objective formulations within the ARJA framework. (Average of 30 runs)}
\label{tab-ms} \footnotesize \tabcolsep =3pt
\begin{tabular}{ccccccccc}
\toprule
\multirow{2}{2.8em}{Bug\\Index}       &   \multicolumn{2}{c}{Success}   &&   \multicolumn{2}{c}{Patch Size}  &&   \multicolumn{2}{c}{\#Patches} \\
\cmidrule{2-3} \cmidrule{5-6} \cmidrule{8-9}
  &              ARJA  & $\text{ARJA}_{s}$      &&  ARJA  & $\text{ARJA}_{s}$  &&  ARJA  & $\text{ARJA}_{s}$    \\                                            
\midrule

F1 & 30 & 26 && 2.00 & 3.04 && 16.50 & 1.73\\
F2 & 17 & 13 && 1.35 & 2.85 && 10.94 & 1.54\\
F3 & 26 & 4 && 2.12 & 3.25 && 4.00 & 1.00\\
F4 & 13 & 10 && 2.23 & 4.50 && 5.92 & 1.40\\
F5 & 30 & 30 && 2.67 & 2.50 && 7.23 & 4.27\\
F6 & 30 & 28 && 2.80 & 3.86 && 9.77 & 1.61\\
F7 & 25 & 11 && 2.24 & 5.91 && 6.00 & 1.27\\
F8 & 29 & 22 && 2.14 & 4.23 && 7.76 & 1.50\\
F9 & 6 & 0 && 2.33 & -- && 2.50 & --\\
F10 & 18 & 11 && 3.00 & 4.45 && 3.22 & 1.09\\
F11 & 20 & 18 && 3.00 & 6.11 && 3.20 & 1.17\\
F12 & 28 & 15 && 3.07 & 4.07 && 6.61 & 1.80\\
F13 & 20 & 15 && 3.15 & 5.47 && 4.60 & 1.27\\
\bottomrule
\end{tabular}
\begin{tablenotes}
        \item ``--'' means the data is not available. 
\end{tablenotes}
\end{threeparttable}
\end{table}

More interestingly, in terms of ``Success'' metric, we find that ARJA also clearly outperforms $\text{ARJA}_{s}$.
Considering that this metric only concerns the weighted failure rate ($f_{2}$ formulated in Eq. (\ref{eq-f2})),
our results suggest that the simultaneous minimization of $f_{1}$ and $f_{2}$ promotes 
the minimization of $f_{2}$ significantly. So, in the sense of search or optimization, 
$f_{1}$ can be seen as a helper objective in our multi-objective formulation of program repair. 
A similar phenomenon was also observed by some previous studies \cite{knowles2001reducing,jensen2004helper,yuan2015multiobjective} on other applications,
which is formally termed as \emph{multi-objectivization} \cite{knowles2001reducing} in the literature. 
One possible reason for this improvement  is that a helper
objective can guide the search toward solutions containing better building blocks
and helps the search to escape local minima \cite{jensen2004helper}.

To sum up, the multi-objective formulation helps
to find simpler repairs (containing smaller number of edits) and also helps to
find more of them. Furthermore, the multi-objective formulation can facilitate
more effective search of test-suite adequate patches compared to the single-objective formulation. 

\subsection{Strength in Fixing Multi-Location Bugs (RQ4)}
\label{sec-Strength in Fixing Multi-Location Bugs (RQ4)}

\begin{table*}[htbp]
\renewcommand{\arraystretch}{1.0}
\centering
\begin{threeparttable}
\caption{Comparison of ARJA, GenProg, and RSRepair on multi-location bugs. (Average of 30 runs)}
\label{tab-sfm} \footnotesize \tabcolsep =4pt
\begin{tabular}{cccccccccccccccccccc}
\toprule
\multirow{2}{2.8em}{Bug\\Index}       &   \multicolumn{3}{c}{Success}               &&   \multicolumn{3}{c}{\#Evaluations}   &&     \multicolumn{3}{c}{CPU (s)} &&     \multicolumn{3}{c}{Patch Size} \\
\cmidrule{2-4} \cmidrule{6-8} \cmidrule{10-12} \cmidrule{14-16} 
                                                          &   ARJA  & GenProg & RSRepair               &&   ARJA & GenProg & RSRepair      &&   ARJA & GenProg & RSRepair  &&   ARJA & GenProg & RSRepair   \\                                
\midrule
F3 & 26 & 23 & 0 && 494.24 & 628.64 & -- && 634.91 & 193.69 & -- && 2.12 & 3.09 & --\\
F4 & 13 & 2 & 0 && 746.54 & 1240.50 & -- && 980.77 & 1129.29 & -- && 2.23 & 6.50 & --\\
F5 & 30 & 11 & 4 && 384.63 & 1235.30 & 1000.50 && 98.33 & 248.73 & 138.80 && 2.67 & 4.10 & 2.00\\
F6 & 30 & 11 & 0 && 624.80 & 894.91 & -- && 393.25 & 127.18 & -- && 2.80 & 7.09 & --\\
F7 & 25 & 4 & 0 && 698.52 & 820.00 & -- && 461.89 & 186.60 & -- && 2.24 & 7.00 & --\\
F8 & 29 & 24 & 3 && 376.21 & 915.21 & 1028.33 && 551.50 & 678.05 & 507.51 && 2.14 & 6.79 & 2.33\\
F9 & 6 & 1 & 0 && 1028.00 & 225.00 & -- && 962.62 & 52.15 & -- && 2.33 & 2.00 & --\\
F10 & 18 & 2 & 0 && 936.11 & 1896.00 & -- && 190.01 & 280.70 & -- && 3.00 & 4.50 & --\\
F11 & 20 & 9 & 0 && 777.70 & 1325.00 & -- && 532.54 & 378.61 & -- && 3.00 & 11.33 & --\\
F12 & 28 & 15 & 0 && 742.04 & 973.73 & -- && 566.57 & 273.59 & -- && 3.07 & 9.67 & --\\
F13 & 20 & 8 & 0 && 762.30 & 1383.00 & -- && 485.07 & 357.65 & -- && 3.15 & 14.88 & --\\

\bottomrule
\end{tabular}
\begin{tablenotes}
        \item ``--'' means the data is not available. 
\end{tablenotes}
\end{threeparttable}
\end{table*}

Most  existing program repair systems (e.g., Nopol) can only generate
single point repairs. GenProg and RSRepair are two 
state-of-the-art repair approaches that can target
multi-location bugs.
To assess the strength of ARJA in multi-location repair, we compare it 
with GenProg and RSRepair on the bugs F3--F13.  F1 and F2 are not considered here
since they belong to single-location bugs. 

Note that ARJA does not take advantage of GenProg and RSRepair when comparing them on
F3--F13, because all  three approaches are based on the redundancy assumption and 
the fix ingredients of F3--F13 exist in their search space. 

Table \ref{tab-sfm} shows the comparative results of ARJA, GenProg and RSRepair on
F3--F13. As can be seen, ARJA outperforms both GenProg and RSRepair on all the
considered bugs in terms of success rate. Indeed, on most of these bugs,
ARJA  achieves a much higher success rate than its counterparts.
Compared to GenProg and RSRepair, ARJA also 
generally requires much smaller number of evaluations to find a repair.
Although GenProg achieves better ``\#Evaluations'' on F9, the metric is computed only
based on a single successful trial. Given that ARJA does more than GenProg and RSRepair in one fitness evaluation,
we also report the results of ``CPU (s)'' to provide another reference for 
comparing the efficiency of the approaches. It can be seen that the 
overall CPU time consumed by ARJA is comparable to that by GenProg. 
Since RSRepair only fixes two bugs here with very low success rate, 
we cannot rate its efficiency. 
In terms of ``Patch Size'', ARJA  usually finds a much simpler
repair  than GenProg. This could be explained by different search
mechanisms of GenProg and ARJA. 
GenProg  considers larger patches in each generation by appending new edits to
the existing patches. So once GenProg cannot
find a repair in the first few generations, it will usually obtain 
patches that contain a relatively high number of edits. Different from GenProg, ARJA prefers smaller
patches throughout the search process. 
In addition, we find that GenProg performs significantly better than
RSRepair on the  multi-location bugs considered, which corroberates our 
speculation from Section \ref{sec-Genetic Search vs. Random Search (RQ2)}.

In summary, ARJA exhibits critical superiority over two prominent repair approaches (i.e., GenProg and RSRepair)
in fixing multi-location bugs.

\subsection{Effect of Type Matching (RQ5)}
\label{sec-Effect of Type Matching (RQ5)}

To show the effect of the type matching strategy, we introduce 
 three additional ARJA variants denoted as $\text{ARJA}_{v}$, 
 $\text{ARJA}_{m}$ and $\text{ARJA}_{b}$. 
They do not use the direct approach for ingredient screening. Instead,
$\text{ARJA}_{v}$ uses the type matching strategy  just for
variables (illustrated in Fig. \ref{fig-vmat}), $\text{ARJA}_{m}$ uses this 
strategy for just methods (illustrated in Fig. \ref{fig-mmat}), and $\text{ARJA}_{b}$
conducts type matching for both variables and methods. 

Table \ref{tab-tm} compares ARJA (without type matching), $\text{ARJA}_{v}$, 
$\text{ARJA}_{m}$ and $\text{ARJA}_{b}$ on bugs H1--H5, where ``Success'' is used
as the comparison metric. From Table \ref{tab-tm} we can see that, 
ARJA cannot find any test-suite adequate patch for all the bugs
except H4, whereas $\text{ARJA}_{v}$ or $\text{ARJA}_{b}$ have a good
chance to fix these bugs. This indicates type matching
is a promising strategy that can help ARJA to fix some 
bugs whose fix ingredients do not exist in the buggy program considered. However, 
$\text{ARJA}_{m}$ does not perform very well here, which may imply that
type matching for methods struggles to generate the
fix ingredients for bugs H1--H3 and H5. Note that ARJA can fix bug H4, 
which means that, in terms of semantics, the repair mode of H4 still follows the redundancy 
assumption.

\begin{table}[htbp]
\renewcommand{\arraystretch}{1.0}
\centering
\caption{Illustration of the type matching strategy (the metric ``Success'' is reported in this table) (30 runs)}
\label{tab-tm} \footnotesize \tabcolsep =5pt
\begin{tabular}{cccccccc}
\toprule
\multirow{2}{2.5em}{Bug\\Index}  &  \multirow{2}{*}{$k$} &  \multirow{2}{*}{$|T_{f}|$} &  \multirow{2}{*}{ARJA}   &  \multirow{2}{*}{$\text{ARJA}_v$}  &  \multirow{2}{*}{$\text{ARJA}_m$}  &  \multirow{2}{*}{$\text{ARJA}_{b}$}\\
       \\
\midrule

H1 & 2 & 6 & 0 & 2 & 0 & 0\\
H2 & 2 & 6 & 0 & 24 & 0 & 20\\
H3 & 2 & 5 & 0 & 4 & 0 & 3\\
H4 & 2 & 2 & 28 & 16 & 25 & 19\\
H5 & 2 & 7 & 0 & 17 & 0 & 13\\
\bottomrule
\end{tabular}
\end{table}

Although $\text{ARJA}_{v}$ and $\text{ARJA}_{b}$ perform much better overall than ARJA
on these bugs, we note that they fail to achieve a very high success rate, particularly on H1 and H3.
Considering that $\text{ARJA}_{v}$ and $\text{ARJA}_{b}$ indeed search over a much larger
space of ingredient statements than ARJA, a possible reason is that the larger search space
poses a serious difficulty for the underlying genetic search. More CPU time may help $\text{ARJA}_{v}$ and $\text{ARJA}_{b}$
to overcome this difficulty. 

To sum up, the type matching strategy shows good potential to generate useful
ingredient statements. These new ingredients can be exploited by the genetic search
to fix  bugs for which the redundancy assumption does not hold. 
However, the much larger search space also challenges the search ability of GP in the
ARJA framework.

\subsection{Evaluation on Real-World Bugs (RQ6)}
\label{sec-Evaluation on Real-World Bugs (RQ6)}

\begin{table*}[htbp]
\renewcommand{\arraystretch}{1.0}
\centering
\begin{threeparttable}
\caption{Results for 224 bugs considered in Defects4J with ten repair approaches. For each approach, we list the bugs for which the test-suite adequate patches are found}
\label{tab-rb} \footnotesize \tabcolsep =7.3pt
\begin{tabular}{clllll}
\toprule
Project & ARJA & $\text{ARJA}_v$ & $\text{ARJA}_m$   & $\text{ARJA}_b$ & GenProg  \\
\midrule
\multirow{4}{*}{JFreeChart} & C1, C3, C5, C7, & C1, C3, C5, C7, & C1, C3, C5, C7, & C1, C3, C7, C12, & C1, C3, C7, C12, \\
& C12, C13, C15, C19, & C12, C13, C15, C25& C12, C13, C15, C18, & C13, C15, C19, C25& C13, C18, C25\\
& C25& & C19, C25& & \\
\cmidrule{2-6}
& $\sum = 9$& $\sum = 8$& $\sum = 10$& $\sum = 8$& $\sum = 7$\\
\midrule
\multirow{3}{*}{JodaTime} & T4, T11, T14, T15& T1, T4, T11& T4, T11, T15, T24& T1, T4, T11, T14, & T4, T11, T24\\
& & & & T17& \\
\cmidrule{2-6}
& $\sum = 4$& $\sum = 3$& $\sum = 4$& $\sum = 5$& $\sum = 3$\\
\midrule
& L7, L16, L20, L22, & L7, L13, L14, L22, & L7, L20, L22, L27, & L7, L13, L14, L21, & L7, L22, L35, L39, \\
& L35, L39, L41, L43, & L35, L39, L43, L45, & L39, L43, L45, L50, & L22, L35, L39, L45, & L43, L51, L55, L59, \\
Commons & L45, L46, L50, L51, & L51, L55, L59, L60, & L51, L55, L58, L59, & L46, L51, L55, L59, & L63\\
Lang& L55, L59, L60, L61, & L63& L60, L61, L63& L60, L61, L63& \\
& L63& & & & \\
\cmidrule{2-6}
& $\sum = 17$& $\sum = 13$& $\sum = 15$& $\sum = 15$& $\sum = 9$\\
\midrule
& M2, M5, M6, M8, & M2, M6, M8, M20, & M2, M5, M6, M8, & M2, M5, M6, M7, & M2, M6, M8, M20, \\
& M20, M22, M28, M31, & M28, M31, M39, M49, & M20, M22, M28, M31, & M8, M20, M28, M49, & M28, M31, M39, M40, \\
& M39, M49, M50, M53, & M50, M53, M56, M70, & M39, M40, M44, M49, & M50, M53, M58, M60, & M49, M50, M71, M73, \\
Commons & M56, M58, M60, M68, & M71, M73, M79, M80, & M50, M53, M58, M70, & M71, M73, M80, M81, & M80, M81, M82, M85, \\
Math& M70, M71, M73, M74, & M81, M82, M85, M95& M71, M73, M74, M79, & M82, M84, M85, M95, & M95\\
& M80, M81, M82, M84, & & M80, M81, M82, M84, & M103& \\
& M85, M86, M95, M98, & & M85, M86, M95, M98, & & \\
& M103& & M103& & \\
\cmidrule{2-6}
& $\sum = 29$& $\sum = 20$& $\sum = 29$& $\sum = 21$& $\sum = 17$\\
\midrule
Total & 59 (26.3\%)& 44 (19.6\%)& 58 (25.9\%)& 49 (21.9\%)& 36 (16.1\%)\\

\toprule
\toprule
Project & RSRepair & Kali & jGenProg\tnote{1}  & jKali\tnote{1} & Nopol\tnote{1}  \\
\midrule

\multirow{3}{*}{JFreeChart}& C1, C5, C7, C12, & C1, C5, C12, C13, & C1, C3, C5, C7, & C1, C5, C13, C15, & C3, C5, C13, C21, \\
& C13, C15, C25& C15, C25, C26& C13, C15, C25& C25, C26& C25, C26\\
\cmidrule{2-6}
& $\sum = 7$& $\sum = 7$& $\sum = 7$& $\sum = 6$& $\sum = 6$\\
\midrule
\multirow{2}{*}{JodaTime}& T4, T11, T24& T4, T11, T24& T4, T11& T4, T11& T11\\
\cmidrule{2-6}
& $\sum = 3$& $\sum = 3$& $\sum = 2$& $\sum = 2$& $\sum = 1$\\
\midrule
& L7, L22, L27, L39, & L7, L22, L27, L39, & & & L39, L44, L46, L51, \\
Commons& L43, L51, L59, L60, & L44, L51, L55, L58, & & & L53, L55, L58\\
Lang& L63& L63& & & \\
\cmidrule{2-6}
& $\sum = 9$& $\sum = 9$& $\sum = 0$& $\sum = 0$& $\sum = 7$\\
\midrule
& M2, M5, M6, M8, & M2, M8, M20, M28, & M2, M5, M8, M28, & M2, M8, M28, M32, & M32, M33, M40, M42, \\
& M18, M20, M28, M31, & M31, M32, M49, M50, & M40, M49, M50, M53, & M40, M49, M50, M78, & M49, M50, M57, M58, \\
& M39, M49, M50, M53, & M80, M81, M82, M84, & M70, M71, M73, M78, & M80, M81, M82, M84, & M69, M71, M73, M78, \\
Commons& M58, M68, M70, M71, & M85, M95& M80, M81, M82, M84, & M85, M95& M80, M81, M82, M85, \\
Math & M73, M74, M80, M81, & & M85, M95& & M87, M88, M97, M104, \\
& M82, M84, M85, M95, & & & & M105\\
& M103& & & & \\
\cmidrule{2-6}
& $\sum = 25$& $\sum = 14$& $\sum = 18$& $\sum = 14$& $\sum = 21$\\
\midrule
Total & 44 (19.6\%)& 33 (14.7\%)& 27 (12.1\%)& 22 (9.8\%)& 35 (15.6\%)\\
\bottomrule
\end{tabular}
\begin{tablenotes}
        \item[1] The results are organized according to those reported in ref. \cite{martinez2016automatic}.
\end{tablenotes}
\end{threeparttable}
\end{table*}

In this subsection, we conduct a large-scale experiment on the Defects4J dataset, in order to
show the superiority of the proposed repair approaches in fixing real-world bugs. 
Our experiment here is similar to that by Martinez et al. \cite{martinez2016automatic} but involves
a larger number of repair approaches. Specifically, 
we consider ARJA along with its three variants (i.e., $\text{ARJA}_v$, $\text{ARJA}_m$ and $\text{ARJA}_b$), GenProg,
RSRepair and Kali, which are all implemented by ourselves under the same infrastructure. Moreover, for our comparison purposes,
we also include the results of jGenProg, jKali and Nopol reported in ref. \cite{martinez2016automatic} on the same dataset.
Note that the time-out for all three approaches was set to three hours per repair attempt. According to the experiments by Martinez et al. \cite{martinez2016automatic},  
a larger time-out would not improve their effectiveness.
The implemented approaches 
generally need less than 1.5 hours to find the first test-suite adequate patch (with very few exceptions) and
we use a CPU environment similar to the one used in ref. \cite{martinez2016automatic}, so the comparison here is unlikely to favor our implemented approaches.

\begin{figure}[htbp]
\centering
\begin{lstlisting}[ linebackgroundcolor={%
  	\ifnum\value{lstnumber}>3
                    \ifnum\value{lstnumber}<8
                \color{green!35}
            \fi
          \fi
        }]
// DateTimeZone.java
public long adjustOffset(long instant, boolean earlierOrLater) {
...
+  instantAfter = 
+  	FieldUtils.safeAdd(instantAfter, 
+		((long) hashCode()) * ((long) 
+			DateTimeConstants.MINUTES_PER_DAY));
   return convertLocalToUTC(local, false, 
  		earlierOrLater ? instantAfter : 
			instantBefore);
}
\end{lstlisting} 
\caption{Test-suite adequate patch generated by $\text{ARJA}_{b}$ for the bug T17.}
\label{fig-T17}
\end{figure}

Table \ref{tab-rb} summarizes the results of the ten repair approaches on 224 bugs considered in Defects4J. 
For each approach, we list every bug for which at least one test-suite adequate patch is found. 
From Table \ref{tab-rb}, ARJA is able to fix the highest number of bugs among all the ten approaches, with a total of 59,
which accounts for 26.3\% of all the bugs considered. To our knowledge, 
none of the existing repair approaches in the literature can synthesize test-suite adequate patches for so many bugs
on the same dataset. 

Although each of the three ARJA variants (using type matching) searches over a 
superset of ARJA's ingredient space, they do not repair a higher number of bugs compared 
to ARJA. The possible reason is that the 
search ability of GP is still not strong enough to handle such a large search space determined by
type matching, which has also been mentioned in Section \ref{sec-Effect of Type Matching (RQ5)}.
However, we note that the three ARJA variants can fix a few bugs that cannot be patched by 
any redundancy-based repair approach (including ARJA) in comparison. These bugs are T1, L13, L14 and M79 by $\text{ARJA}_v$;
L58, M44 and M79 by $\text{ARJA}_m$; and T1, T17, L13, L14, L21 and M7 by $\text{ARJA}_b$.
This demonstrates the effectiveness of type matching on some real-world bugs. For instance, only
$\text{ARJA}_{b}$ synthesizes a test-suite adequate patch for T17, which is shown in Fig. \ref{fig-T17}.
The  statement inserted (lines 4--7 in Fig. \ref{fig-T17}) indeed does not exist in the buggy program considered, whereas the
following statement does
\begin{lstlisting}[numbers=none,frame =none]
minutes = FieldUtils.safeAdd(minutes, ((long) getDays()) * ((long) DateTimeConstants.MINUTES_PER_DAY));
\end{lstlisting} 
But the variable \lstinline[language=Java]!minutes! and the method \lstinline[language=Java]!getDays()! in this statement
are both outside the scope of the faulty location. $\text{ARJA}_{b}$ maps the variable
\lstinline[language=Java]!minutes! to \lstinline[language=Java]!instantAfter! and the method invocation
\lstinline[language=Java]!getDays()! to  \lstinline[language=Java]!hashCode()! through type matching, thereby 
inventing a new statement. $\text{ARJA}_{b}$ exploits GP to insert this new statement before line 8, which allows
the patched program to fulfill the given test suite.

\begin{figure}[htbp]
\centering
\subfloat[ARJA, GenProg, RSRepair and Kali]{\includegraphics[width=0.45\columnwidth]{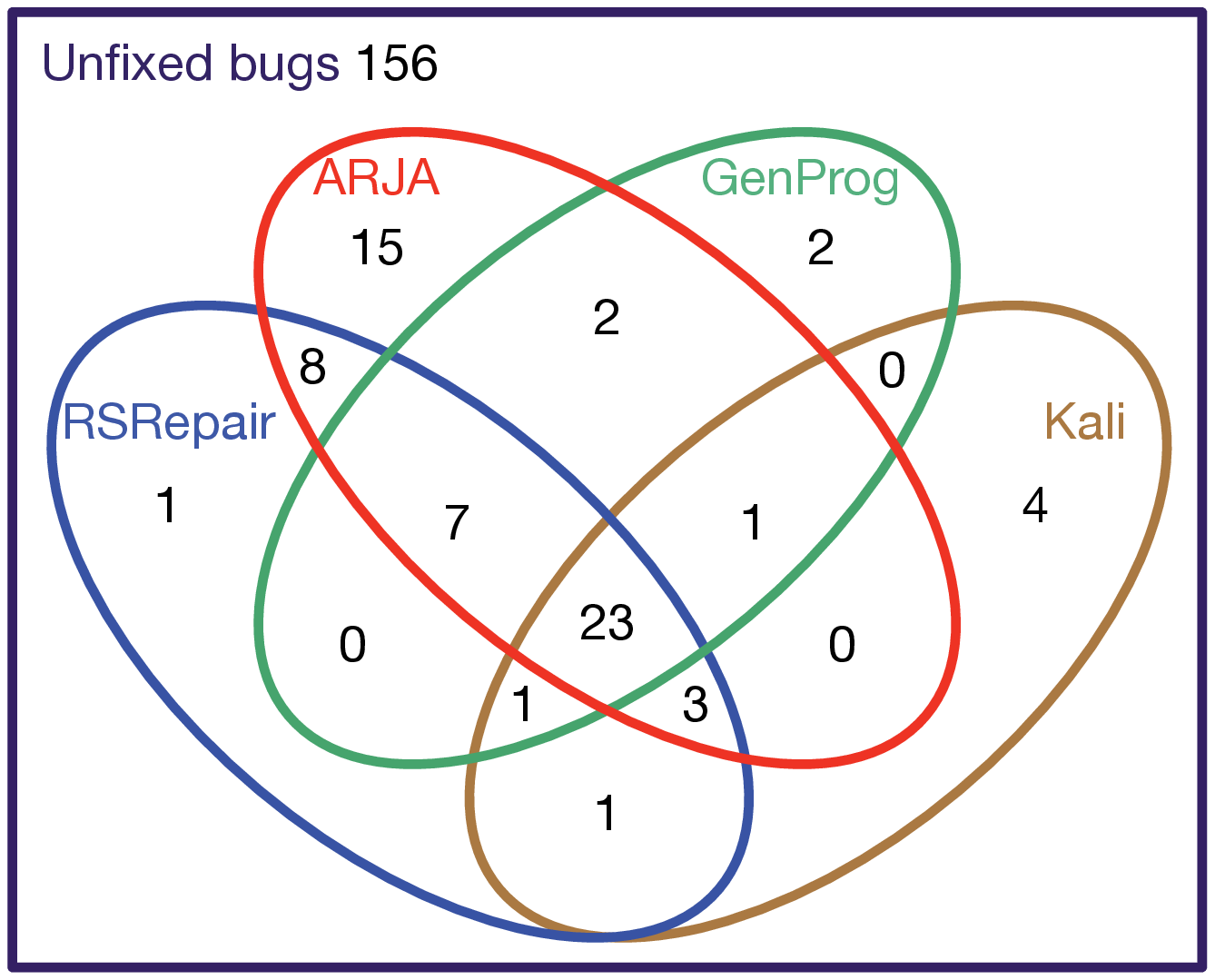}%
\label{fig-intersect1}}
\hspace{20pt}
\subfloat[ARJA, jGenProg, jKali and Nopol]{\includegraphics[width=0.45\columnwidth]{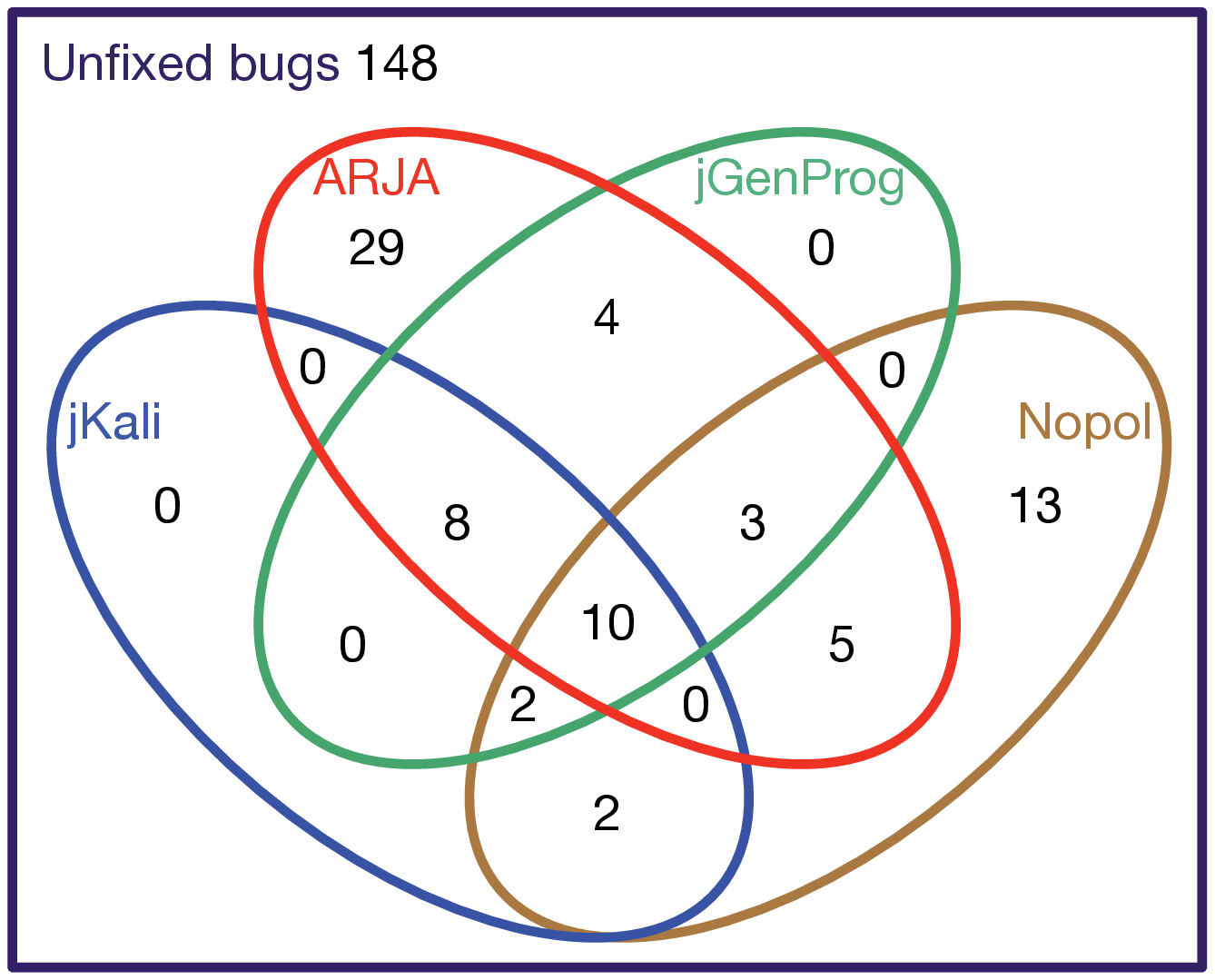}%
\label{fig-intersect2}}
\caption{Venn diagram of bugs for which test-suite adequate patches are found.}
\label{fig-inter}
\end{figure}

Another three repair approaches implemented by ourselves (i.e., GenProg, RSRepair and Kali) fix 36, 44 and 33 bugs respectively.
To show their performance difference with ARJA more clearly, Fig. \ref{fig-inter}\subref{fig-intersect1}
presents a Venn diagram that shows the intersections of the fixed bugs among the four approaches.  
As seen from Fig. \ref{fig-inter}\subref{fig-intersect1},
the overwhelming majority of bugs handled by GenProg, RSRepair and Kali can also be handled by ARJA;
ARJA is able to fix 15 bugs that neither GenProg, RSRepair nor Kali could fix;
for 23 bugs, all the four repair approaches can generate a test-suite adequate patch.
Note that on the Defects4J dataset considered, 
RSRepair can synthesize test-suite adequate patches for more bugs than GenProg. Further, we find that
RSRepair generates a patch containing only a single edit for 41 out of 44 fixed bugs.
Recall that the search mechanism of RSRepair is more suitable to find
such very simple repairs than that of GenProg, as discussed in Section \ref{sec-Genetic Search vs. Random Search (RQ2)}.
So, it is not surprising that RSRepair shows a certain advantage over GenProg on the bugs considered here.

jGenProg, jKali and Nopol can synthesize test-suite adequate patches for 27, 22 and 35 bugs
respectively. Fig. \ref{fig-inter}\subref{fig-intersect2} compares ARJA with the three approaches
using the Venn diagram. ARJA is clearly superior to jGenProg and jKali. Almost all the bugs repaired by jGenProg and jKali can also
be repaired by ARJA. ARJA also fixes a greater number of bugs than Nopol, but their performance shows good complementarity:
ARJA and Nopol can handle 41 and 17 bugs, respectively, that cannot be handled by the peer. 

In addition, our implemented GenProg and Kali show better overall performance than jGenProg or jKali. 
The reason may be due to different parameter settings and implementation methods. 

To summarize, ARJA and its three variants have an obvious advantage over the other state-of-the-art repair approaches
in handling real-world bugs. We find that the ten repair approaches under consideration in our experiment can 
synthesize a test-suite adequate patch for 88 out of total 224 bugs. To provide a reference for future research, the patches generated by 
our  approaches are publicly available at GitHub.\footnote{Defects4J patches, http://github.com/yyxhdy/defects4j-patches}

\subsection{Patch Correctness (RQ7)}
\label{sec-Patch Correctness (RQ7)}

Because of a weak test suite used as  oracle, a test-suite adequate patch for
certain bugs may be incorrect though passing the given test suite, a condition called patch overfitting \cite{qi2015analysis,smith2015cure}. 
In this subsection, we manually evaluate the correctness of the patches generated by ARJA.\footnote{Due to limited manual effort, we currently do not consider the correctness of the patches found by the other approaches in comparison}
We regard a test-suite adequate patch correct if it is exactly the same as or semantically equivalent 
to a human-written patch. Due to a lack of domain knowledge, we cannot confirm the correctness of the test-suite adequate patches for some bugs, and do not include them in our manual assessment.

 \begin{table}[htbp]
\renewcommand{\arraystretch}{1.0}
\centering
\caption{The bugs for which the correct patches are synthesized by ARJA}
\label{tab-correct} \footnotesize \tabcolsep =12pt
\begin{tabular}{cl}
\toprule
Project & Bug ID \\
\midrule
JFreeChart           &  C3, C5, C12\\
\cmidrule{1-2}
Joda-Time            & T15\\
\cmidrule{1-2}
Commons Lang   & L20, L35, L43, L45\\
\cmidrule{1-2}
\multirow{2}{*}{Commons Math}   & M5, M22, M39, M50, M53,  \\
			   &	M58, M70, M73, M86, M98 \\
\midrule
Total     &         18\\
\bottomrule
\end{tabular}
\end{table}

After careful analysis, we find that ARJA can synthesize  correct patches for at least 18 bugs in Defects4J, which are 
shown in Table \ref{tab-correct}. These results are very encouraging. This is because ARJA referred to here is
also based on the redundancy assumption like jGenProg, but jGenProg can only correctly fix 5 bugs (as reported in \cite{martinez2016automatic}) which
are also correctly repaired by ARJA. Among the remaining 13 bugs, jGenProg cannot even find a test-suite adequate patch for 11 of them. 
This again demonstrates the effectiveness of the improved GP search in ARJA. 
Furthermore, to our knowledge, only ARJA can generate a correct patch for  bugs C12, L20, M22, M39, M58, M86 
and M98, whereas the other repair systems in the literature cannot. Another highlight of 
ARJA is that it can correctly fix some multi-location bugs in Defects4J, which are hard to  repair
by the other repair methods.

To illustrate the expressive power, we conduct the
case studies of the bugs that are correctly repaired by ARJA. We find that 
some of these repairs appear to be very interesting.

\subsubsection{Case Study of Single-Location Bugs that are\\ Correctly Repaired by ARJA}
\label{sec-Case Study of Single-Location Bugs that are Correctly Repaired by ARJA}

Among the bugs correctly fixed by ARJA, 13  can be categorized as single-location bugs since
ARJA is able to repair them with only a single edit. Here we only take M58 and M86 as examples.

Fig. \ref{fig-M58} shows the correct patch generated by ARJA for M58. It is syntactically different from the human-written
patch that replaces the faulty statement (lines 6--7) with \lstinline[language=Java]!return fit(guess);!. Nevertheless,
the method parameter in the statement inserted by ARJA (lines 8--10) is indeed equivalent to the variable \lstinline[language=Java]!guess! according
to the variable declaration statement (lines 3--5), and we have confirmed that the method invocations 
 \lstinline[language=Java]!ParameterGuesser!, \lstinline[language=Java]!getObservations! and \lstinline[language=Java]!guess! do not
change anything outside the faulty method \lstinline[language=Java]!fit()!. Thus the patch shown in Fig. \ref{fig-M58}
is semantically equivalent to the human-written patch. 

\begin{figure}[htbp]
\begin{lstlisting}[ linebackgroundcolor={%
  	\ifnum\value{lstnumber}>5
                    \ifnum\value{lstnumber}<8
                \color{red!35}
            \fi
          \fi
    	\ifnum\value{lstnumber}>7
                    \ifnum\value{lstnumber}<11
                \color{green!35}
            \fi
          \fi
        }]
// GaussianFitter.java        
public double[] fit() {
   final double[] guess = (new 
   	ParameterGuesser(getObservations()))
		  .guess();
-  return fit(new Gaussian.Parametric(), 
-		   guess);
+  return fit((new 
+	   ParameterGuesser(getObservations()))
+		  .guess());
}\end{lstlisting}%
\caption{Correct patch generated by ARJA for bug M58.}
\label{fig-M58}
\end{figure}

\begin{figure}[htbp]
\begin{lstlisting}[ linebackgroundcolor={%
  	    	\ifnum\value{lstnumber}>13
                    \ifnum\value{lstnumber}<19
                \color{green!35}
            	   \fi
                 \fi
        }]
// CholeskyDecompositionImpl.java
public CholeskyDecompositionImpl(...)  {
  for (int i = 0; i < order; ++i) {
    final double[] lI = lTData[i];
    if (lTData[i][i] < 
    	absolutePositivityThreshold) {
      throw new 
        NotPositiveDefiniteMatrixException();
    }
    ...
  }
  for (int i = 0; i < order; ++i) {
    final double[] ltI = lTData[i]; 
+   if (lTData[i][i] < 
+     absolutePositivityThreshold) {
+     throw new 
+       NotPositiveDefiniteMatrixException();
+   }
    ...	
  }
 ...
}
\end{lstlisting}
\caption{Correct patch generated by ARJA for the bug M86.}
\label{fig-M86}
\end{figure}

In Fig. \ref{fig-M86}, we show the correct patch found by ARJA for the bug M86. This bug occurs
because the buggy program fails to correctly check whether a symmetric matrix is positive definite (the Cholesky decomposition only applies to the positive-definite matrix).
The buggy program does such a check using the \lstinline[language=Java]!if! statement at line 5 in Fig. \ref{fig-M86}, 
which examines whether all diagonal elements are positive. However this is only a necessary, though not sufficient, condition for 
the positive definite matrix. 
The human-written patch first deletes the \lstinline[language=Java]!if! statement at line 5 and then inserts almost the same \lstinline[language=Java]!if! statement
(\lstinline[language=Java]!lTData[i][i]! is equivalently changed to \lstinline[language=Java]!ltI[i]!) 
before line 19, so that the validation of the positive definitiveness is conducted correctly during the decomposition process. Unlike the human-written patch, 
the correct patch by ARJA does not delete the \lstinline[language=Java]!if! statement at line 5.
Because this \lstinline[language=Java]!if! statement states a necessary condition, just keeping it intact
would not influence the correctness of the patched program.

\subsubsection{Case Study of Multi-Location Bugs that are\\ Correctly Repaired by ARJA}
\label{sec-Case Study of Multi-Location Bugs that are Correctly Repaired by ARJA}

The bugs L20, L35, T15, M22 and M98 are classified as multi-location bugs, since
ARJA fixes each of them correctly using more than one edit. For M22 and M98, ARJA can 
generate a correct patch that is exactly the same as the human-written patch. 
As for the remaining three, ARJA synthesizes semantically equivalent patches, which are
analyzed as follows. 
\begin{figure}[htbp]
\begin{lstlisting}[ linebackgroundcolor={%
  	\ifnum\value{lstnumber}>3
                    \ifnum\value{lstnumber}<8
                \color{red!35}
            \fi
          \fi
            	\ifnum\value{lstnumber}>7
                    \ifnum\value{lstnumber}<10
                \color{green!35}
            \fi
          \fi
                      	\ifnum\value{lstnumber}>13
                    \ifnum\value{lstnumber}<19
                \color{red!35}
            \fi
          \fi
                      	\ifnum\value{lstnumber}>18
                    \ifnum\value{lstnumber}<21
                \color{green!35}
            \fi
          \fi
        }]
// StringUtils.java
public static String join(Object[] array, char separator, int startIndex, int endIndex) { 
  ...
-  StringBuilder buf = new 
-  	 StringBuilder((array[startIndex] == 
-		    null ? 16 : array[startIndex]
-			    .toString().length()) + 1);
+  StringBuilder buf = new 
+       StringBuilder(256);	
  ...		
}
public static String join(Object[] array, String separator, int startIndex, int endIndex) { 
  ...
-  StringBuilder buf = new 
-  	 StringBuilder((array[startIndex] == 
-		    null ? 16 : array[startIndex]
-			    .toString().length()) + 
-				   separator.length());
+  StringBuilder buf = new 
+       StringBuilder(256);	
  ...		
}
\end{lstlisting}%
\caption{Correct patch generated by ARJA for  bug L20.}
\label{fig-L20}
\end{figure}
In Fig. \ref{fig-L20}, we present a correct patch synthesized by ARJA for  bug L20. The reason leading to this bug is that
even if \lstinline[language=Java]!array[startIndex]! is not equal to \lstinline[language=Java]!null!, \lstinline[language=Java]!array[startIndex].toString()!
can still be \lstinline[language=Java]!null!, and \lstinline[language=Java]!array[startIndex].toString().length()! would thereby cause the \lstinline[language=Java]!NullPointerException!. 
The human developer fixes this bug by replacing two faulty statements (lines 4--7 and lines 14--18 in Fig. \ref{fig-L20}) with the same statement shown as follows:
\begin{lstlisting}[numbers=none,frame =none]
StringBuilder buf = new StringBuilder(noOfItems * 16);
\end{lstlisting}
We find that the initial capacity (e.g., \lstinline[language=Java]!noOfItems * 16! or 256) of \lstinline[language=Java]!StringBuilder! has nothing to 
do with the correctness but with the performance. If the the initial capacity is too large, much unnecessary 
memory will be allocated; if it is too small, \lstinline[language=Java]!StringBuilder! will frequently expand its capacity when accommodating additions, 
requiring more computational time. But in terms of making the buggy program functionally correct, the patch generated by ARJA has no difference
with the human-written patch.

Fig. \ref{fig-L35} shows the correct patch found by ARJA for  bug L35. The buggy program fails to satisfy a desired behavior:
the method \lstinline[language=Java]!add(T[] array, T element)! should throw \lstinline[language=Java]!IllegalArgumentException! when both
parameters are \lstinline[language=Java]!null!. The only difference between the patch by ARJA and 
the human-written patch lies in the detailed message of the exception. However, this difference will not 
affect the ability of the patched program by ARJA to meet the specified functionality successfully.

\begin{figure}[htbp]
\begin{lstlisting}[ linebackgroundcolor={%
            	\ifnum\value{lstnumber}>4
                    \ifnum\value{lstnumber}<7
                \color{green!35}
            \fi
          \fi
        \ifnum\value{lstnumber}>11
                    \ifnum\value{lstnumber}<14
                \color{green!35}
            \fi
          \fi
             \ifnum\value{lstnumber}=4
                \color{red!35}
        \fi
   \ifnum\value{lstnumber}=11
                \color{red!35}
        \fi
        }]
// ArrayUtils.java
public static <T> T[] add(T[] array, T element) {
...
-   type = Object.class;
+   throw new IllegalArgumentException
+	   ("The Integer did not match any ...");
...
}
public static <T> T[] add(T[] array, int index, T element) {
...
-   return (T[]) new Object[] { null };
+   throw new IllegalArgumentException
+	   ("The Integer did not match any ...");
...
}
\end{lstlisting}%
\caption{Correct patch generated by ARJA for  bug L35.}
\label{fig-L35}
\end{figure}

ARJA fixes  bug T15 correctly in an interesting way as shown in Fig. \ref{fig-T15}.
This bug occurs when \lstinline[language=Java]!val1! and \lstinline[language=Java]!val2! are equal to \lstinline[language=Java]!Long.MIN_VALUE! and -1 respectively.
In this scenario, the product should be \lstinline[language=Java]!-Long.MIN_VALUE!. But 
\lstinline[language=Java]!-Long.MIN_VALUE! exceeds the maximum allowed value (i.e., \lstinline[language=Java]!Long.MAX_VALUE!) and
the buggy program indeed returns an incorrect value due to overflow. To fix this bug, 
the human developer just inserts the following \lstinline[language=Java]!if! statement before line 5 in Fig. \ref{fig-T15}:  
\begin{lstlisting}[numbers=none,frame =none]
if (val1 == Long.MIN_VALUE) {
   throw new ArithmeticException("...overflows"); 
}
\end{lstlisting}
So in term of the human-written patch, this bug can also be regarded as a single-location bug. 
However, ARJA cannot generate such a patch since the above \lstinline[language=Java]!if! statement
does not exist anywhere in the buggy program. 
As shown in Fig. \ref{fig-T15}, the patch by ARJA first replaces line 5 with a \lstinline[language=Java]!break!
statement to avoid returning an incorrect value there, then it detects the overflow that triggers the bug
in the new \lstinline[language=Java]!if! statement (lines 15--20) with the expression \lstinline[language=Java]!val1 == Long.MIN_VALUE && val2 == -1!.
Note that the boolean expression \lstinline[language=Java]!val2 == Long.MIN_VALUE && val1 == -1! is always false since 
\lstinline[language=Java]!val2! is an \lstinline[language=Java]!int! value, so it has no effect and can be ignored. As can be seen,
the patch by ARJA just does the same thing as the human-written patch in a different way, and thus it is correct. 

\begin{figure}[htbp]
\begin{lstlisting}[ linebackgroundcolor={%
             \ifnum\value{lstnumber}=5
                \color{red!35}
        \fi
   \ifnum\value{lstnumber}=6
                \color{green!35}
        \fi
          \ifnum\value{lstnumber}>10
                    \ifnum\value{lstnumber}<15
                \color{red!35}
            \fi
          \fi
  \ifnum\value{lstnumber}>14
                    \ifnum\value{lstnumber}<21
                \color{green!35}
            \fi
          \fi
        }]
// FieldUtils.java
public static long safeMultiply(long val1, int val2) {
   switch (val2) {
   case -1:
-  return -val1;
+  break;
   case 0: return 0L;
   case 1: return val1;
   }
   long total = val1 * val2;
-  if (total / val2 != val1) {
-     throw new 
-     	ArithmeticException("...overflows");
-  }
+  if (total / val2 != val1 || val1 == 
+    Long.MIN_VALUE && val2 == -1 || val2 == 
+	    Long.MIN_VALUE && val1 == -1)  {
+     throw new 
+  	  ArithmeticException("...overflows");
+  }
   return total;
}
\end{lstlisting}%
\caption{Correct patch generated by ARJA for  bug T15.}
\label{fig-T15}
\end{figure}

\subsubsection{Other Findings}
\label{sec-Other Findings}

Our manual study also provides  other findings besides the
correct patches for some bugs. 

We find that although the test-suite adequate patches for a number of bugs (e.g., C1, C19, L7 and L16) may not be 
correct, they present some similarities  with  corresponding human-written patches. So these
test-suite adequate patches would still be useful in assisting the human developer to 
create correct patches. 

With stronger search ability, ARJA can identify more under-specified bugs than 
previous repair approaches (e.g., jGenProg and jKali). For example,
Martinez et al. \cite{martinez2016automatic}
claimed that the specification of the bug L55 by the test suite is good enough to drive the generation of a correct patch, considering Nopol can repair this bug whereas jGenProg and jKali cannot.
However we find that L55 is also an under-specified bug and an overfitting patch (shown in Fig. \ref{fig-L55})
that simply deletes two statements can fulfill its test suite. 

\begin{figure}[htbp]
\centering
\begin{lstlisting}[ linebackgroundcolor={%
        \ifnum\value{lstnumber}=4 
                \color{red!35}
        \fi
       \ifnum\value{lstnumber}=5
                \color{red!35}
        \fi
        }]
// StopWatch.java
public void stop() {
...
-   stopTime = System.currentTimeMillis();
-   this.runningState = STATE_STOPPED;
}
\end{lstlisting} 
\caption{Test-suite adequate but incorect patch for  bug L55.}
\label{fig-L55}
\end{figure}

Moreover, we have also checked the correctness of the patches by ARJA for
 seeded bugs. We find that most of these test-suite adequate patches are correct. 
Recalling that our implemented Kali cannot uncover
the weakness of the test suite for any seeded bug, this implies that 
a stronger test suite would render ARJA  more able to generate
correct patches. Several recent studies \cite{yu2017test,xin2017identifying,yang2017better} have started to
explore the potential of test case augmentation for program repair.

\subsubsection{Summary}
\label{sec-Summary}

In summary, through careful manual assessment, we find  that ARJA can synthesize
a correct patch for at least 18 bugs in Defects4J. To the best of
our knowledge, some of the 18 bugs have never been fixed correctly by
 existing repair systems in the literature. Furthermore, ARJA is able to 
generate correct patches for several multi-location bugs, which is impossible 
for most of the other repair approaches. 

Note that we do not focus on 
the number of correctly repaired bugs on the Defects4J dataset when comparing
ARJA with  other approaches that are \emph{not} based on the redundancy assumption (e.g., Nopol). 
Nowadays, it is common knowledge that different kinds of repair techniques 
can be better at addressing different classes of bugs. For example, although Nopol that targets conditional 
statement bugs can only fix 5 bugs correctly on the same dataset \cite{martinez2016automatic}, 3 of them cannot
be repaired correctly by ARJA. So the number of correct fixes by different categories of 
repair techniques would strongly depend on how the dataset tested was built \cite{monperrus2014critical}. 
Also, we cannot expect that the 224 bugs in Defects4J can truly reflect the natural distribution of 
real-world bugs. For instance, Defects4J indeed contains a considerable number of null pointer bugs, so
it may favor those approaches that can explicitly conduct null pointer detection with fix templates (e.g., PAR \cite{kim2013automatic} and ACS \cite{xiong2017precise}). 
Compared to such approaches, ARJA is a more generic repair approach.

In contrast, ARJA and jGenProg can be compared meaningfully on the Defects4J dataset in terms of the number of bugs fixed or correctly fixed,
because both of them typically belong to redundancy-based repair techniques
and use GP to explore the search space. ARJA performs much better than jGenProg, which clearly validates the improvement of GP search in ARJA. 

To facilitate re-examination by the other researchers, we provide a detailed explanation of the correctness for each correct patch generated by ARJA,  publicly available at GitHub.\footnote{Correctness, http://github.com/yyxhdy/defects4j-correctness}

\subsection{Reasons for Failure (RQ8)}
\label{Reasons for Failure (RQ8)}

 As seen from the experimental results,
ARJA and its variants sometimes fail to find a test-suite adequate
patch for some bugs. 
We find that there are three possible reasons for  failure, which are discussed in the following.

The first reason is that 
 fix ingredients for the bug do not exist in
the search space of the proposed repair approach. 
In this case, no matter how powerful the 
underlying genetic search is, ARJA (or its variants)
will definitely fail to fix the bug. The failure of ARJA on H1--H3 and H5 (see Table \ref{tab-tm}) can 
be attributed to this reason. 

The second reason is that 
although  test-suite adequate (or even correct) patches
exist in the search space, the search ability of GP
is still not strong enough to find it within the required number of generations. An example is
that ARJA fails on bug F9 in 24 out of 30 trials. Another example is
the failure of ARJA on the real-world bug L53. 
Fig. \ref{fig-L53} shows a human-written patch for this bug. 
We find that ARJA takes into account lines 11 and 19 as potential faulty 
lines by fault localization. So, a correct patch within the search space of ARJA consists of the following two edits:
1) insert the \lstinline[language=Java]!if! statement located at line 7 before line 11; 2) insert the \lstinline[language=Java]!if! statement located at line 15 
before line 19. This patch is semantically equivalent to the human-written patch shown in Fig. \ref{fig-L53}.
However, we note ARJA fails to find it under the parameter settings of our experiment. This may be because
the genetic search is easily trapped in local optima  when navigating the search space corresponding to L53. 

\begin{figure}[htbp]
\centering
\begin{lstlisting}[ linebackgroundcolor={%
        \ifnum\value{lstnumber}=6 
                \color{green!35}
        \fi
       \ifnum\value{lstnumber}=14
                \color{green!35}
        \fi
        \ifnum\value{lstnumber}=10
                \color{red!35}
        \fi
        \ifnum\value{lstnumber}=18
                \color{red!35}
        \fi
        }]
// DateUtils.java
private static void modify(Calendar val, int field, boolean round) {
  ...
  if (!round || millisecs < 500) {
    time = time - millisecs;
+ }
    if (field == Calendar.SECOND) {
      done = true; 
    }
- }
  int seconds = val.get(Calendar.SECOND);
  if (!done && (!round || seconds < 30)) {
    time = time - (seconds * 1000L);
+ }
    if (field == Calendar.MINUTE) {
      done = true;
    }
- }
  int minutes = val.get(Calendar.MINUTE);
  ...
}
\end{lstlisting} 
\caption{Human-written patch for  bug L53.}
\label{fig-L53}
\end{figure}
 
The last reason is that ARJA fails to consider the faulty 
lines that trigger the bug in to computational search, which can be further divided into the following three categories:
\begin{enumerate}
\item The fault localization technique adopted in ARJA fails to identify all  faulty lines related to the bug of interest. This applies to
 bugs C2, T9, L4, M12, M104, and so on. 
\item Due to the inadequate accuracy of fault localization, the faulty lines are given relatively low suspiciousness. 
For example, Fig. \ref{fig-L10} shows the human-written patch for L10. We find that the suspiciousness for
all these faulty lines is less than 0.2, but the number of lines with suspiciousness larger than 0.2 is more than 40 (i.e., $n_{\max}$ value in our experiments).
Hence ARJA leaves out  faulty lines and fails to fix this bug. 
\item The test suite is not adequate for the fault localization. Bug M46 provides an example of such a scenario, whose human written
patch is shown in Fig. \ref{fig-M46}. We find that the whole method starting from line 10 is not covered by any negative test. 
So based on the current test suite of M46, the coverage-based fault localization technique, no matter how powerful, cannot identify line 13 as a potential faulty line. 
\end{enumerate}

\begin{figure}[htbp]
\centering
\begin{lstlisting}[ linebackgroundcolor={%
        \ifnum\value{lstnumber}=3 
                \color{red!35}
        \fi
        \ifnum\value{lstnumber}>4
                    \ifnum\value{lstnumber}<13
                \color{red!35}
            \fi
          \fi
           }]
// FastDateParser.java
private static StringBuilder escapeRegex(StringBuilder regex,...) {
- boolean wasWhite = false;
    ...
-   if(Character.isWhitespace(c)) {
-     if(!wasWhite) {
-       wasWhite = true;
-       regex.append("\\s*+");
-     }
-     continue;
-   }
-   wasWhite = false;
    ...
}
\end{lstlisting} 
\caption{Human-written patch for  bug L10.}
\label{fig-L10}
\end{figure}
\begin{figure}[htbp]
\centering
\begin{lstlisting}[ linebackgroundcolor={%
        \ifnum\value{lstnumber}=5 
                \color{red!35}
        \fi
       \ifnum\value{lstnumber}=6
                \color{green!35}
        \fi
        \ifnum\value{lstnumber}=13
                \color{red!35}
        \fi
        \ifnum\value{lstnumber}=14
                \color{green!35}
        \fi
        }]
// Complex.java
public Complex divide(Complex divisor) throws NullArgumentException {
  ...
  if (divisor.isZero) {
-   return isZero ? NaN : INF;
+   return NaN;
  }
  ...
}
public Complex divide(double divisor) {
  ...
  if (divisor == 0d) {
-   return isZero ? NaN : INF;
+   return NaN;
  }
  ...
}
\end{lstlisting} 
\caption{Human-written patch for  bug M46.}
\label{fig-M46}
\end{figure}
 
Note that we can use simple strategies to alleviate the issues mentioned in
the second and the third categories. 
For example, for the bug L10, we just reset the parameter $n_{\max}$
to 80 and then run ARJA again. As a result, ARJA can now
find a test-suite adequate patch for L10 which simply deletes the
\lstinline[language=Java]!if! statement at line 5 in Fig. \ref{fig-L10}. 
This patch is semantically equivalent to the human written patch and is
thus correct. 
As for M46, we modify the JUnit test named \lstinline[language=Java]!testScalarDivideZero!
as shown in Fig. \ref{fig-M46_test}. This JUnit test (before modification)
is originally a positive test, because
\lstinline[language=Java]!x.divide(Complex.ZERO)!
and \lstinline[language=Java]!x.divide(0)! return 
the same incorrect value (i.e., \lstinline[language=Java]!INF!) due to 
the faults in lines 5 and 13 in Fig. \ref{fig-M46}, and because
this test only checks the equality rather than the individual values. 
We add a statement (lines 4--5 in Fig. \ref{fig-M46_test})
that can expose the fault at line 13 in Fig. \ref{fig-M46}, and run ARJA again on M46 with the modified test suite. 
As a result, ARJA can now fix this multi-location bug and the patch obtained is exactly the same as the human-written patch. 

\begin{figure}[htbp]
\centering
\begin{lstlisting}[ linebackgroundcolor={%
        \ifnum\value{lstnumber}=4 
                \color{green!35}
        \fi
       \ifnum\value{lstnumber}=5 
                \color{green!35}
        \fi
           }]
// ComplexTest.java
public void testScalarDivideZero() {
   Complex x = new Complex(1,1);
+ TestUtils.assertEquals(x.divide(0), 
+      Complex.NaN, 0);  
   TestUtils.assertEquals(x.divide(
   	Complex.ZERO), x.divide(0), 0);  
}
\end{lstlisting} 
\caption{The modification of an associated JUnit test of M46.}
\label{fig-M46_test}
\end{figure}

\subsection{Repair Efficiency (RQ9)}
\label{sec-Repair Efficiency (RQ9)}
For industrial applicability of automated program repair, an approach should fix a bug within
a reasonable amount of time. Table \ref{tab-repe} presents  time  (in minutes) of generating the first repair
for the Defects4J bugs.
Note that although jGenProg, jKali and Nopol were executed by Martinez et al. \cite{martinez2016automatic} on a  machine different from ours,
we can roughly compare their time cost with that of our implementation since we use a similar CPU environment. 

\begin{table}[htbp]
\renewcommand{\arraystretch}{1.3}
\centering
\begin{threeparttable}
\caption{Time cost of patch generation on Defects4J dataset}
\label{tab-repe} \footnotesize \tabcolsep =8.9pt
\begin{tabular}{lccccc}
\toprule
\multirow{2}{2.8em}{Repair\\Approach}       &   \multicolumn{4}{c}{CPU Time (in minutes)}     \\
         \cmidrule{2-5}    
                  &  Min  &  Median   & Max  & Average    \\
\midrule 

Arja\tnote{1}& 0.73 & 4.70 & 63.73 & 10.02\\
$\text{Arja}_v$\tnote{1} & 0.86 & 4.63 & 80.63 & 10.32\\
$\text{Arja}_m$\tnote{1} & 0.86 & 4.67 & 37.85 & 10.49\\
$\text{Arja}_b$\tnote{1} & 0.89 & 5.27 & 95.12 & 11.48\\
GenProg\tnote{1} & 0.61 & 8.43 & 83.06 & 16.20\\
RSRepair\tnote{1} & 0.87 & 6.23 & 238.93 & 17.88\\
Kali\tnote{1} & 0.89 & 2.38 & 48.05 & 6.58\\

jGenProg\tnote{2}  &  0.67   & 61 & 76 & 55.83 \\

jKali\tnote{2}          & 0.6      &   18.75 & 87 & 23.55  \\

Nopol\tnote{2} & 0.52 & 22.5 & 114 &  30.88 \\
\bottomrule
\end{tabular}
\begin{tablenotes}
         \item[1] The CPU time on an Intel Xeon E5-2680 V4 2.4 GHz processor with 20 GB memory.
        \item[2] The CPU time on an Intel Xeon X3440 2.53 GHz processor with 15 GB memory. The results are excerpted from \cite{martinez2016automatic}.
\end{tablenotes}
\end{threeparttable}
\end{table}

As seen from Table \ref{tab-repe}, the median  and  average time for a successful
repair by ARJA and its three variants runs for around 5 minutes and 10 minutes, respectively.
However, maximum CPU time can reach more than one hour. 
GenProg, RSRepair and Nopol are less efficient than ARJA and its variants, which further
verifies the superiority of the ARJA framework. Kali is the most efficient on average but it only considers
trivial patches. In addition, our implementation of GenProg and Kali show clear advantage over jGenProg and jKali in terms of
time cost possibly attributable to the test filtering procedure or different implementation methods.

In summary, the repair efficiency of ARJA and its variants  compares favorably with 
that of  existing notable automatic repair approaches. Considering that ARJA consumes
about 10 minutes on average and one hour at most for a repair in our experiments, 
we think that this efficiency is generally acceptable for industrial use 
in light of the bug-fixing time required by human programmers \cite{kim2006long,weiss2007long}.

\section{Threats to Validity}
\label{sec-Threats to Validity}
In this section, we discuss three basic types of threats that can affect the 
validity of our empirical studies. 

\subsection{Internal Validity}
\label{sec-Internal Validity}

To support a more reasonable comparison, we reimplemented GenProg, RSRepair and Kali
for Java under the same infrastructure as ARJA. Although we faithfully 
and carefully followed the details described in the corresponding research papers during reimplementation, our implementation 
may still not perform as
well as the original systems. There may even be bugs in the implemented systems that we 
have not found yet. To mitigate this threat, we make the code of
the three systems available on GitHub for peer-review. Note that although jGenProg 
has been widely used as the the implementation of GenProg for Java \cite{le2016history, xiong2017precise, yu2017test, xin2017identifying}, 
it was not implemented by the original authors of GenProg, thereby also potentially suffering from reimplementation issues.
According to our results, our implemented GenProg indeed shows advantages over jGenProg. 

In our experiments, we use the same parameter setting for all  bugs considered. 
There is a risk that the parameter setting is  poor for handling some bugs. 
Section \ref{Reasons for Failure (RQ8)} has shown an example where  resetting of $n_{\max}$ can allow ARJA
find a correct patch for  bug L10. However it is not realistic to select an ideal parameter setting for every repair approach on
every bug. We here use a uniform parameter setting  among tghe implemented repair approaches to ensure
a fair comparison.  

Another internal threat to validity concerns the stochastic nature of ARJA (including its variants), GenProg and RSRepair. 
It is possible that different runs of these repair approaches would obtain somewhat different results. We run
each of them only once on each of 224 bugs in Defects4J, which may lead to an overestimation or underestimation of
their repair effectiveness on this dataset. However, our experiments on the Defects4J dataset have already 
been much larger in scale than those conducted by Martinez et al. \cite{martinez2016automatic}, since it involved a larger
number of repair methods (i.e., 7 methods) and repair trials (i.e., 1,568 trials).

\subsection{External Validity}
\label{sec-External Validity}
Our experimental results are based on both seeded bugs and real-world bugs. 

Although the seeded bugs F1--F15 are  randomly 
produced, there still exists the possibility that the 
fitness landscapes corresponding to these bugs favor 
a certain kind of search mechanism. So the evaluation on
these bugs may not reflect the actual difference in search ability between different 
search strategies (i.e., multi-objective GP, single-objective GP and random search).

For real-world bugs, we used 224 bugs of four Java projects from Defects4J. 
However, it is not possible to expect that such a number of bugs can fully represent 
the actual distribution of defect classes and difficulties in the real world.
So the evaluation may not be adequate enough to reflect the actual effectiveness of 
our repair techniques on real-world bugs, and our results may also not
generalize to other datasets. However, Defects4J is known to be the most comprehensive 
dataset of real Java bugs currently available, and it has been extensively used
as the benchmark for evaluating Java 
program repair systems \cite{martinez2016automatic, le2016history, xiong2017precise, yu2017test, xin2017identifying}.

\subsection{Construct Validity}
\label{sec-Construct Validity}

Here we manually analyze the correctness of
the test-suite adequate patches generated by ARJA. Such a manual study is not scientifically sound, although it has been an accepted practice \cite{qi2015analysis,martinez2016automatic} in automated program repair.
It may happen that the analysts classify an incorrect patch as correct due to a limited understanding
of the buggy program. For example, Martinez et al. \cite{martinez2016automatic} claimed that Nopol can synthesize correct patches for  bugs C5 and M50,
but a recent study \cite{yu2017test} indicated that the generated patches are not really correct. 
To reduce this threat, we carefully rechecked the correctness of the identified correct patches and 
made the explanation,. why we believe they are correct, available online. 


ARJA finishes a repair trial for a bug only when the maximum number of generations
is reached. So ARJA may finally return multiple test-suite adequate patches (with the same patch size) for a bug. 
In such a case, we examine the correctness of all the patches obtained, and we deem ARJA to be
able to fix this bug correctly if at least one of these patches is identified as correct.  
We confirm that ARJA outputs both correct and incorrect (i.e., only test-suite adequate) patches
for  bugs C12, L43 and M22. So a strict criterion for correctness would pose a threat to the validity of the number of correct repairs by ARJA. 
However, we think it is unrealistic to completely avoid the generation of incorrect but test-suite adequate patches when
 program repair is just based on a weak test suite. Possibly  machine learning techniques \cite{long2016automatic} can be used to estimate the probability of a patch being correct, but
this falls outside the scope of this paper. Further, sometimes it is indeed very difficult to differentiate the 
incorrect patches from correct ones. For example, Fig. 27 shows a test-suite adequate patch found by ARJA for  bug L43. 
This patch is indeed incorrect but it is very similar to the human-written patch that inserts \lstinline[language=Java]!next(pos);! before line 6 rather than line 5. 
Even an experienced human programmer cannot easily recognize it as incorrect without a deep understanding of the program specification.
We still think test-suite augmentation is a fundamental solution to the patch overfitting problem. 
\begin{figure}[htbp]
\centering
\begin{lstlisting}[ linebackgroundcolor={%
        \ifnum\value{lstnumber}=4 
                \color{green!35}
        \fi
           }]
// ExtendedMessageFormat.java
 private StringBuffer appendQuotedString(...) {
...
+ next(pos);
   if (escapingOn && c[start] == QUOTE) {
     return appendTo == null ? null : 
          appendTo.append(QUOTE);
   }
}
\end{lstlisting} 
\caption{Test-suite adequate patch found by ARJA for  bug L43.}
\label{fig-L43}
\end{figure}

Different from ARJA, jGenProg is terminated once the first test-suite adequate patch is found, and 
the analysis of correctness only targets this patch. This may lead to a bias toward ARJA
when comparing its number of correct fixes with that of jGenProg. 
jGenProg may still have had the chance to find a correct patch for certain bugs, had it not been
terminated immediately. However, we think such bias is minimal: Except  for
5 bugs correctly fixed by both ARJA and jGenProg, jGenProg could not produce any patch for 11 out of the remaining 13
bugs that are correctly repaired by ARJA. 

\section{Related Work}
\label{sec-Related Work}

The test-suite based program repair techniques can be 
roughly divided into two main branches: search-based repair approaches and 
semantics-based repair approaches. In this section, we first list related studies 
about these two categories of approaches, then review the research on
 empirical aspects of test-suite based repair.

\subsection{Search-Based Repair Approaches}
\label{sec-Search-Based Repair Approaches}

Search-based repair approaches generally determine a search space that 
potentially contains  correct repairs, and then apply computational search techniques, 
such as a GA or random search, to navigate the search space, in order to
find  test-suite adequate patches.

\textbf{JAFF.} Arcuri and Yao \cite{arcuri2008novel} proposed the idea 
of using GP to co-evolve the programs and unit tests, in order to fix a
bug automatically. Subsequently, Arcuri \cite{arcuri2011evolutionary}  
developed a research prototype, called JAFF, which models bug fixing
as a search problem and uses EAs to solve it. JAFF can only handle a 
subset of Java and was evaluated on toy programs.

\textbf{GenProg.} GenProg is a prominent GP based program repair system which was developed
jointly by several researchers \cite{weimer2009automatically,forrest2009genetic,le2012genprog,le2012systematic}. 
Le Goues et al. \cite{le2012systematic} presented the latest GenProg implementation, where the patch 
representation is used instead of the AST based representation \cite{weimer2009automatically,forrest2009genetic,le2012genprog} 
in order to make GenProg scalable to large-scale programs. It was reported in \cite{le2012systematic}
that GenProg can automatically repair 55 out of 105 bugs from 8 open-source programs. Since our study is
based on GenProg, a more detailed description was given in Section \ref{sec-Brief Introduction to GenProg}.

Around the GenProg framework, a number of related studies have been conducted in the literature. 
Fast et al. \cite{fast2010designing} investigated two approaches (i.e., test suite sampling and dynamic predicates) to enhancing fitness functions in GenProg.
Schulte et al. \cite{schulte2010automated} applied GenPog to fix bugs in x86 assembly and Java bytecode programs.
Le Goues et al. \cite{le2012representations} investigated the choices of solution representation and genetic operators 
for the underlying GP in GenProg. 
Oliveira et al. \cite{oliveira2016improved} presented a low-granularity patch representation and developed
several crossover operators associated with this representation. 
Tan et al. \cite{tan2016anti} suggested a set of anti-patterns to inhibit GenProg or the other search-based methods 
from generating nonsensical patches.

\textbf{Mutation-based repair.} Debroy and Wong \cite{debroy2010using}
proposed to combine standard mutation operators (from the mutation testing literature) and fault
localization to fix bugs. This method considers each possibly faulty location one by one 
according to the suspiciousness metric, and mutates the statement at the current location to 
produce potential fixes.

\textbf{PAR.} Kim et al. \cite{kim2013automatic} proposed PAR, which leverages 
fix patterns manually learned from human written patches. 
Similar to GenProg, PAR also implements an evolutionary computing process.
But instead of using crossover and mutation operators as in GenProg, PAR uses
fix templates derived from common fix patterns to produce new program variants in each generation. 
Experiments on six Java projects and a user study confirm that the patches generated by PAR are
often more meaningful than those by GenProg. 

\textbf{AE.} Weimer et al. \cite{weimer2013leveraging} proposed a
deterministic repair algorithm based on program equivalence, called AE.
This algorithm uses adaptive search strategies
to control the order in which candidate repairs and test cases are considered. 
Empirical evaluations showed that, AE can reduce the search space by an
order of magnitude when compared to GenProg.

\textbf{RSRepair.} Qi et al. \cite{qi2014strength} presented RSRepair, which replaces
the evolutionary search in GenProg with random search. Their experiments on 24 bugs of the 
GenProg benchmark suite
indicate that random search performs more effectively and efficiently than GP 
in program repair. 

\textbf{SPR.}  Long and Rinard \cite{long2015staged} reported SPR, which adopts
a staged program repair strategy to navigate a rich search space of candidate patches efficiently. 
SPR defines a set of transformation schemas beforehand and uses the target value search or 
condition synthesis algorithm to determine the parameter values of the selected transformation schema.
Experimental results on 69 bugs from 8 open source applications indicate that SPR
can generate more correct patches than  previous repair systems.

\textbf{Prophet.}  Based on SPR, Long and Rinard \cite{long2016automatic}
further designed Prophet, a repair system using
a probabilistic model to rank  candidate patches
in the search space of SPR. Given a training set of successful human patches, Prophet learns model parameters via maximum
likelihood estimation. Experimental results indicate that a learned model 
in Prophet can significantly improve its ability to generate correct patches.

\textbf{HistoricalFix.} Le et al. \cite{le2016history} introduced a repair method which 
evolves  patches based on bug fix patterns mined from the history of many projects.
This method uses 12 kinds of existing mutation operators to generate candidate patches and determines
the fitness of a patch by assessing its similarity to the mined bug-fixing patches. Experimental results on 90 bugs from Defects4J show
that the proposed method can produce good-quality fixes for more bugs compared to GenProg and PAR.

\textbf{ACS.}  Xiong et al. \cite{xiong2017precise} reported ACS, a repair system targeting
\lstinline[language=Java]!if! condition bugs. ACS decomposes the condition synthesis
into two steps: variable selection and predicate selection. Based on the decomposition,  
it uses  dependency-based ordering along with the information of 
javadoc comments to rank the variables, and uses predicate mining to rank predicates.
With a synthesized condition, ACS leverages three fix templates to generate a patch.
Experiments show that ACS can successfully repair 18 bugs from four projects of Defects4J.

\textbf{Genesis.} Long et al. \cite{long2017automatic} presented 
a repair system, called Genesis, which can automatically infer code transforms and
search spaces for patch generation. Genesis was tested on two classes of errors (i.e., null pointer errors and out of bounds errors) in
real-world Java programs.

\subsection{Semantics-Based Repair Approaches}
\label{sec-Semantics-Based Repair Approaches}

Typically, the semantics-based repair approaches infer semantic constraints 
from the given test cases, and then generate the test-suite adequate patch through solving the resulting 
constraint satisfaction problem, particularly the SMT problem. 

\textbf{SemFix.} Nguyen et al. \cite{nguyen2013semfix} proposed SemFix, 
a pioneering tool for semantic-based program repair. SemFix first employs 
statistical fault localization to identify likely-buggy statements. Then, for each
identified statement, it generates repair constraints through symbolic execution of the given test suite
and solves the resulting constraints by an SMT solver. SemFix targets  faulty locations that 
are either a right hand side of an assignment or a branch
predicate, and was compared to GP based methods on programs
with seeded  as well as real bugs.

\textbf{SearchRepair.} Ke et al. \cite{ke2015repairing} developed  
a repair method based on semantic code search, called SearchRepair. 
This method encodes a database of human-written code fragments as SMT constraints on
the input-output behavior and searches the database for potential fixes with an input-output specification.
Experiments on small C programs written by students showed that SearchRepair can generate higher-quality 
repairs than GenProg, AE and RSRepair.

\textbf{DirectFix.} Mechtaev et al. \cite{mechtaev2015directfix} implemented a prototype system, called DirectFix, for automatic program repair.
To consider the simplicity of repairs, 
DirectFix integrates fault localization and patch generation into a single step by leveraging partial maximum SMT constraint solving and 
component-based program synthesis. Experimental comparison indicates that the patches found by 
DirectFix are simpler and safer than those by SemFix.

\textbf{QLOSE.} D'Antoni et al. \cite{d2016qlose} formulated the quantitative program repair problem,
where the optimal repair is obtained by minimizing an objective function combing several 
syntactic and semantic distances to the buggy program.
The problem is an instance of maximum SMT problem and is solved by an existing program synthesizer.
The technique was implemented in a prototype tool called QLOSE and was evaluated on programs taken from educational tools.

\textbf{Angelix, JFix.} Mechtaev et al. \cite{mechtaev2016angelix} presented Angelix, a semantic-based repair method
that is more scalable than SemFix and DirectFix. The scalability of Angelix attributes to
the new lightweight repair constraint (called an angelic forest), which is independent of the size 
of the program under repair. 
Experimental studies on the GenProg benchmark indicate that Angelix has the ability to fix bugs from large-scale software and multi-location bugs.
Angelix was originally designed for C programs. Recently, Le et al. \cite{le2017jfix} developed JFix which is an extension of
Angelix for Java.

\textbf{Nopol.} Xuan et al. \cite{xuan2017nopol} proposed an approach, called Nopol,
for automatic repair of buggy conditional statements in Java programs. 
Nopol employs angelic fix localization to identify potential fix locations and expected values of \lstinline[language=Java]!if! conditions. 
For each identified location,
it encodes the test execution traces as a SMT problem and converts
the solution to this SMT into a patch for the buggy program. Nopol was
evaluated on 22 bugs in real-world programs.

\textbf{S3.} Le et al. \cite{le2017s3} presented S3 which leverages the programming-by-examples methodology
to synthesize high-quality patches. S3 was evaluated on 52 bugs in small programs and 100 bugs
in real-world large programs, and experimental results show that it can
generate more high-quality repairs than several existing semantic-based repair methods.

\subsection{Empirical Aspects of Test-Suite Based Repair}
\label{sec-Empirical Aspects of Test-Suite Based Repair}

Besides proposing new program repair methods, there is another line of research that focuses
on the empirical aspects of test-suite based repair.

\textbf{Patch maintainability.}  
Fry et al. \cite{fry2012human} presented a human study of patch maintainability involving
150 participants and 32 real-world defects. Their results indicate that 
machine-generated patches are slightly less maintainable than human-written ones. 
Tao et al. \cite{tao2014automatically} investigated an application scenario of
automatic program repair where the auto-generated patches
are used to aid the debugging process by humans.

\textbf{Redundancy assumption.} Martinez et al. \cite{martinez2014fix} investigated
all commits of 6 open-source Java projects experimentally, and found that a large
portion of commits can be composed of what has already existed in
previous commits, thereby validating the fundamental redundancy assumption of GenProg. 
In the same year, Barr et al. \cite{barr2014plastic} inquired whether the
redundancy assumption (or plastic surgery hypothesis) holds by examining 15,723 commits
from 12 large Java projects. Their results show that 43\% changes can 
be reconstituted from existing code, thus promising success to  the repair methods
that search for fix ingredients in the  buggy program considered.

\textbf{Patch overfitting.}  
Qi et al. \cite{qi2015analysis} analyzed the patches reported by three existing patch generation
systems (i.e., GenProg, RSRepair, and AE), and found that most of these  are not
correct and are equivalent to a single functionality deletion,
due to either the use of weak proxies or weak test suites.
Based on this observation, they  presented Kali which generates
patches only by deleting functionality. Their experiments show that
Kali can find at least as many plausible patches than three prior systems on the
GenProg benchmark. 
Smith et al. \cite{smith2015cure} conducted a controlled empirical study of
GenProg and RSRepair on the IntroClass benchmark. 
By using two test suites for each program (one for patch generation and another for evaluation),
their experiments identified the circumstances under which  patch overfitting
happens.  
Long and Rinard. \cite{long2016analysis} analyzed the search spaces for 
patch generation systems. Their analysis indicates that  correct patches occur sparsely
within the search spaces and that plausible patches are relatively abundant compared to
correct patches. They suggest using information
other than the test suite to isolate correct patches. 
Yu et al. \cite{yu2017test} investigated
the feasibility and effectiveness of test case generation in
addressing the overfitting problem. Their results indicate that test case generation 
is ineffective at promoting the generation of correct patches, thus calling for 
research on test case generation techniques tailored to program repair systems. 
Xin et al. \cite{xin2017identifying} proposed a tool named DiffTGen, which could
identify overfitting patches through test case generation. They also showed
that a repair method configured with DiffTGen could avoid obtaining overfitting
patches and potentially generate correct ones. 
Yang et al. \cite{yang2017better} presented an overfitting patch detection framework which can 
filter out overfitting patches by enhancing existing test cases.

\textbf{Analysis of real-world bug fixes.} 
Martinez and Monperrus \cite{martinez2015mining} proposed to mine
repair actions from software repositories. Based on
a fine-grain AST differencing tool, their work analyzed 62,179 versioning transactions
extracted from 14 repositories of open-source Java software,
in order to obtain the probability distributions over different repair actions. 
It was expected that such distributions can guide the search of 
repair methods.  
Zhong and Su \cite{zhong2015empirical} conducted a large-scale empirical investigation 
on over 9,000 bug fixes from 6 popular Java projects, then distilled several findings and
insights that can help improve state-of-the-art repair methods. 
Soto et al. \cite{soto2016deeper} presented a large-scale empirical study of 
bug fix commits in Java projects. Their work provided several insights about 
broad characteristics, fix patterns, and statement-level mutations in real-world bug fixes,
motivating additional study of repair for Java. 

\textbf{Performance evaluation.}
Kong et al. \cite{kong2015experience} compared four program repair techniques on 
153 bugs from 9 small to medium sized programs, and investigated the impacts of
different programs and test suites on effectiveness and efficiency of the techniques in comparison.
Martinez et al. \cite{martinez2016automatic} conducted a large-scale empirical evaluation of program
repair methods on 224 bugs in Defects4J. Their experimental results showed that three considered
methods (i.e., GenProg, Kali, and Nopol) can generate test-adequate patches for 47 bugs, among which
9 bugs were confirmed to be repaired correctly by manual.
Le et al. \cite{le2016empirical} presented an empirical comparison of different synthesis engines for
semantics-based repair approaches on IntroClass benchmark. 
Durieux et al. \cite{durieux2017patches} reported the test-adequate patches obtained by Nopol on
the bugs of Defects4J version 1.1.0. 

\textbf{Influence of fault localization.}  
Qi et al. \cite{qi2013using} evaluated the effectiveness 
of 15 popular fault localization techniques when plugged
into GenPog. Their work claims that automated fault localization
techniques need to be studied from the viewpoint of fully automated debugging.
Assiri and Bieman \cite{assiri2017fault} experimentally evaluated the impact of 10 fault location
techniques on the effectiveness, performance, and repair correctness of a brute-force repair method. 
Wen et al. \cite{wen2017empirical} conducted controlled experiments using the Defects4J dataset
to investigate the influence of the fault space on a
typical search-based repair approach (i.e., GenProg).

\textbf{Datasets.}  
Just et al. \cite{just2014defects4j} presented Defects4J which is a bug database containing 357 real bugs from 5 real-world open-source Java projects.
Le Goues et al. \cite{le2015manybugs} designed two datasets (i.e., ManyBugs and IntroClass), which consist of
1,183 bugs in 15 C programs and support the comparative evaluation of repair algorithms for 
various of experimental questions.
Tan et al. \cite{tan2017codeflaws} presented a dataset, called Codeflaws, where
all  3,902 defects contained from 7,436 C programs are classified into 39 defect classes.
Berlin \cite{berlin2017bug} collected a dataset called DBGBench, which consists of
27 real bugs in widely-used C programs and can serve as reality check for debugging and
repair approaches.

\section{Conclusion and Future Work}
\label{sec-Conclusion and Future Work}

In this paper, we have proposed ARJA, a new GP based program repair
approach for Java. 
Specifically, we present a lower-granularity patch representation which properly decouples the 
search subspaces of likely-buggy locations, operation types and ingredient statements, thereby
enabling GP to traverse the search space more effectively. 
Based on this new representation, we propose to view automated program repair
as a multi-objective optimization problem of 
minimizing the weighted failure rate and patch size simultaneously, and then use
a multi-objective GA (i.e., NSGA-II) to search for simpler repairs. 
To speed up the fitness evaluation of GP, we present a test 
filtering procedure that can ignore a number of tests that 
are unrelated to the GP manipulations. Considering the characteristic of Java,
we propose to restrict the replacement/insertion code to that which are not only
in the variable scope but also in the method scope at the destination, so that
the modified program is more likely to be compiled successfully.
To further reduce the search space, we design three
types of rules that are seamlessly integrated into three different phases (i.e., operation initialization, ingredient screening and solution decoding)
of ARJA. In addition, we present a type matching strategy that can
exploit the syntactic patterns of the statements that are out of scope at the destination
so as to invent some new ingredient statements that are potentially useful. The type matching
strategy can be optionally integrated into ARJA. 

We conduct a large-scale experimental study on both seeded bugs and real-world bugs. 
The evaluation on seeded bugs clearly demonstrates the necessity and effectiveness of multi-objective GP used in ARJA, 
and also illustrates the strength of the type matching strategy. Furthermore, we evaluate
ARJA and its three variants using type matching 
on 224 real-world bugs from Defects4J, in comparison 
with several state-of-the-art repair approaches. The comparison results show that 
ARJA can generate a test-suite adequate patch for the highest 
number (i.e., 59) of real bugs, as opposed to only 27 by jGenProg and 35 by Nopol. 
The three ARJA variants can fix some bugs that cannot be fixed by ARJA, 
showing the potential of type matching on real-world bugs. Manual 
analysis confirms that ARJA can correctly fix 18 bugs at least in Defects4J, as opposed to
5 by jGenProg. To our knowledge, there are 7 among the 18 bugs that are repaired automatically and correctly 
for the first time. Another highlight is that ARJA can correctly repair several multi-location bugs that are 
widely recognized as hard to be repaired. Our study strongly suggests that the power of GP
for program repair was far from being fully exploited in the past, and that GP can be expected to perform much better
on this important and challenging task. 

ARJA is publicly available at GitHub to facilitate further reproducible research on automated Java program repair: 
\url{http://github.com/yyxhdy/arja}.

In the future, we plan to incorporate a number of repair templates \cite{kim2013automatic} into our ARJA framework so as to further
enhance its performance on real-world bugs. Moreover, considering the mutational robustness \cite{schulte2014software} in software,
we would like to combine the infrastructure of ARJA
and the advanced many-objective GAs \cite{yuan2016new,yuan2016balancing} to improve the non-functional properties \cite{white2011evolutionary,langdon2015optimizing} of Java software. 




\end{document}